\documentclass{aastex63}
\usepackage{amsmath}
\usepackage{mathtools}
\submitjournal{ApJ}

\shorttitle{Chaos in Binaries}
\shortauthors{Seligman $\&$ Batygin}
\graphicspath{{./}{figures/}}

\begin{document}

\title{The Onset of Chaos in Permanently Deformed Binaries from Spin-Orbit and Spin-Spin Coupling}

\correspondingauthor{Darryl Seligman}
\email{dzseligman@uchicago.edu}

\author{Darryl Seligman}
\affiliation{ Dept. of the Geophysical Sciences, University of Chicago, Chicago, IL 60637}

\author{Konstantin Batygin}
\affil{Division of Geological and Planetary Sciences, Caltech, Pasadena, CA 91125}




\begin{abstract}
Permanently deformed objects in binary systems can experience complex rotation evolution, arising from the extensively studied effect of spin-orbit coupling as well as more nuanced dynamics arising from spin-spin interactions. The ability of an object to sustain an aspheroidal shape largely determines whether or not it will exhibit non-trivial rotational behavior. In this work, we  adopt a simplified model of a gravitationally interacting primary and satellite pair, where each body's quadrupole moment is approximated by two diametrically opposed point masses. After calculating the net gravitational torque on the satellite from the primary, and the associated equations of motion, we  employ a Hamiltonian formalism which allows for a perturbative treatment of the spin-orbit and retrograde and prograde spin-spin coupling  states. By analyzing the resonances individually and collectively, we determine the criteria for resonance overlap and the onset of chaos, as a function of orbital and geometric properties of the binary.  We extend the 2D planar geometry to calculate the obliquity evolution, and find that satellites in  spin-spin resonances undergo precession when inclined out of the plane, but do not tumble. We apply our resonance overlap criteria to the contact binary system (216) Kleopatra, and find that its satellites, Cleoselene and Alexhelios, may plausibly be exhibiting chaotic rotational dynamics from the overlap of the spin-orbit and retrograde spin-spin resonances. While this model is by construction generalizable to any binary system, it will be particularly useful to study small bodies in the Solar System, whose irregular shapes make them ideal candidates for exotic rotational states. 
\end{abstract}

\keywords{Astrodynamics --Asteroids--Minor planets--Trans-Neptunian objects}


\section{Introduction} \label{sec:intro}

The  study of spin-axis dynamics  dates back to a simple question: why is the Moon always facing the same side to the Earth? With this simple problem in mind, \citet{Darwin1879,Darwin1880} demonstrated from basic physical principles that the spin-orbit synchronization of a satellite is a natural consequence of dissipation within a fluid orbiting body. They demonstrated that tidal dissipation  caused both the spin-down of the Earth and the outward migration of the Moon. The natural conclusion from this theory is that the Moon and the Earth used to be much closer together and that the pair were once   the same body. The theory of spin-axis dynamics  was extended by \citet{Kaula1964} and \citet{MacDonald1964}. \citet{MacDonald1964} presented a detailed history of the past orbital evolution of the Earth-Moon system, and speculated on the spin states of some of the other planets in the Solar System.  After these pioneering works, spin-synchronization was generally accepted as the only expected outcome of long term tidal evolution.

Radio measurements of Mercury's rotation, however, demonstrated that the geosynchronous state was not the only possible outcome of tidal evolution. Specifically, \citet{Pettengill1965} used radar observations from the Arecibo Ionospheric Observatory to determine that the spin period of Mercury was $\sim$59 days. This was in direct contradiction to the previous belief, since Mercury's orbital period is 88 days. Following this discovery, several theoretical studies  demonstrated that when a body can sustain a permanent quadrupole moment, gravitational torques exerted upon the orbiting object can yield rich dynamical behavior, with a progressively intricate spectrum of resonances emerging with increasing orbital ellipticity.  In a short communication, \citet{Peale1965}  claimed that Mercury's proximity to the Sun implied that it must be in the end state of some dissipative process. They argued that due to the Mercury's moderate eccentricity, its spin-rate should lie between its orbital period and $2\pi$ divided by the orbital velocity at periapse, instead of in the geosynchronous spin-state, as long as its quadrupole moment was small.  It was subsequently discovered that,  owing to its moderate eccentricity, the 3:2 spin-orbit resonance of Mercury constitutes one of the possible outcomes of long term tidal evolution by \citet{Liu1965} and \citet{Colombo1965} independently. \citet{Colombo1965} gave a physical explanation for why the 3:2 spin orbit coupling state is stable, with the axis of minimum moment of inertia aligned with the sun at periastron passage. In a seminal study, \citet{Goldreich1966:spinorbit} demonstrated that resonant states existed at every half integer ratio of the mean motion, and showed that the geosynchronous Lunar motion and 3:2 Mercurian rotation were natural and expected end states of spin-orbit coupling. This theory was further developed in \citet{Goldreich1968}.

The theoretical advances that explained  the discovery of Mercury's 3:2 spin-orbit resonance also helped to explain the tidal evolution of many of the natural satellites in the Solar System. A primary example of this is the  detailed investigations of the violent Earth-Moon history, after the impact generating the Moon \citep{Canup2001,Cameron1976}, with the incorporation of tidal dissipation. Intriguingly, \citet{Makarov2013} pointed out that the Moon could have easily ended in the 3:2 spin-state like Mercury, and argued that the current geosynchronous state was evidence for an initial retrograde rotation or low eccentricity orbit. \citet{Touma1994} presented a generalized scenario including several different models of tidal dissipation for the Earth-Moon history, and calculated  complex dynamical histories of  Earth's obliquity under the action of the evolution of the Moon. \citet{Cuk2012} proposed a formation scenario where an impact formed a large circum-planetary disk, where the Moon formed with significantly higher angular momentum,  and was captured into the current geosynchronous state as a consequence of passage through the corotation resonance, followed by long-term tidal de-spinning.  \citet{Aharonson2012} constrained the quality factor of the Moon, using the current orientation of the Moon's moment of inertia. 


If the permanent distortion of the orbiting body is not merely present, but is sufficiently large, spin-orbit resonances can overlap, giving rise to even richer, chaotic behavior. After pointing this out, \citet{Wisdom1984}  predicted that the Saturnian satellite Hyperion was  chaotically tumbling, because of tidal evolution. This prediction was dramatically confirmed by \citet{Black1995}, who produced clearly irregular light curves of Hyperion using Voyager 2 images.

Chaotic rotation and tumbling can significantly alter the orbital evolution and efficiency of tidal dissipation of a satellite \citep{Wisdom1987}. \citet{Marcialis1987} hypothesized that the geologically active topographical features on the Uranian satellite Miranda could be a natural consequence of chaotic tumbling, similar to that proposed for Hyperion. Since  tidal dissipation is more efficient during chaotic tumbling,  this could have caused the more efficient melting of Miranda. \citet{Dermott1988} argued that this increased efficiency of tidal dissipation would have damped Miranda's eccentricity and may have disrupted previous resonant configurations with the other Uranian satellites. The satellites of the Pluto-Charon binary system \citep{Weaver2006}, Nyx and Hydra, were also believed to be undergoing chaotic rotation due to instability from spin-orbit resonances \citep{Showalter2015}. However, this may not be the case, since New Horizons observations demonstrated that the spin rates of the satellites were much higher than previously assumed, so the tidal spin down timescale is much longer than previously considered \citep{Weaver2016}.

While the rotational dynamics of the planetary moons in the Solar System  historically drove the theoretical advances in the study of tidal evolution and spin dynamics, a large fraction of the minor body populations are also binary.  For detailed reviews of these populations, we refer the reader to \citet{Richardson2006}, \citet{Walsh2009}, \citet{Margot2015}, and Chapters 9-10 in \citet{Prialnik2020}. The first binary minor planet discovered, following the definition provided in \citet{Richardson2006}, was Pluto's moon, Charon \citep{Christy1978}. Early indirect evidence for binary minor bodies was inferred by the fact that $\sim 10 \%$ of the impact structures on the Earth are doublets \citep{Melosh1991}. \citet{Bottke1996} proposed that these doublet craters could be explained by a, then undetected, population of binary near-Earth asteroids, consisting of $\sim 15 \%$ of the population.  Following the serendipitous discovery of Dactyl, the satellite of Ida, by the Gallilleo spacecraft \citep{Chapman1995,Belton1995,Belton1996}, a wealth of information has been garnered about the binary population of Solar System objects. In 1998, a second binary asteroid was discovered, 45 Eugenia, and its companion, Petit Prince \citep{Merline1999}. Early estimates of the binary population of near-Earth objects suggested that $\sim 16 \%$ of large $>200 \rm{m}$ objects were binary \citep{Margot2002}. \citet{Pravec2006} provided a photometric survey of near-Earth asteroids searching for binaries, and found that $\sim 15 \%$ of the population were binaries, confirming the previously predicted population by \citet{Bottke1996}. \citet{Noll2006} discovered the first binary companion to a Centaur, or a giant planet crossing minor body, with the Hubble Space Telescope, which suggested that binaries were common in that population. \citet{Stephens2006} presented detections of six trans-Neptunian object (TNO) binaries in the cold classical population of the Kuiper belt, and concluded that   $\sim 11 \%$ of the TNOs were binary. Curiously, they found that $\sim 22 \%$ of the dynamically cold population were binary, while only $\sim 5 \%$ of the dynamically hot population were binary, a population with significantly higher inclinations \citep{Levison2001,Brown2001}. These estimates were updated by \citet{Noll2008}, who estimated that the binary fractions were  $\sim 30 \%$ and $\sim 10 \%$ for the dynamically cold and hot objects, respectively. 

The formation of these binary Solar System bodies has been an area of rich theoretical investigation. A promising formation mechanism for small near-Earth and main belt binaries is the Yarkovsky-O'Keefe-Radzievskii-Paddack (YORP) effect \citep{Opik1951,Radzievskii1952}, where radiative torques lead to spin-up to the point of rotational fission \citep{Scheeres2007,Walsh2008}. It is also possible that rotational fission could account for a fraction of the trans-Neptunian objects, including the dwarf planet Haumea \citep{Ortiz2012}.  \citet{Goldreich2002} proposed that these binaries formed via collisionless processes, where transient binaries entering each other's Hill sphere's lose energy either due to dynamical friction or gravitational interaction with a third body \citep{Funato2004,Schlichting2008}. \citet{Canup2005} presented hydrodynamical simulations that provided evidence for a giant impact formation of the Pluto-Charon system. Debris from this collision has been postulated to have provided the  material for the satellites of the Pluto-Charon binaries \citep{Stern2006}. In the outer Solar System, one possible route to the formation of binaries is fission during gravitational collapse \citep{Youdin2002,Johansen2007}, perhaps triggered by the streaming instability \citep{Youdin2005}, where excess angular momentum prevents the collapsing object from forming a single body \citep{Nesvorny2010}. Recent hydrodynamical simulations of gravitational collapse via the streaming instability match  the inclination distribution of Kuiper Belt binaries, providing strong evidence that this is a primary formation mechanisms \citep{Nesvorny2019}. Observations of the heterogeneity of the color, reflectance spectra and surface composition of trans-Neptunian binary pairs also supports this formation mechanism \citep{Marsset2020}. However, there do exist a population of blue and wider-separated trans-Neptunian binaries that have been argued to have formed closer in and pushed out \citep{Fraser2017}, although color has been shown to be a notoriously difficult diagnostic of a parent population \citep{Peixinho2015}.

Small bodies that are bound by tensile strength and not-self gravity often have irregular shapes, such as suggested by the modelling in \citet{Lacerda2007}. In a survey presented by \citet{Marchis2006}, a substantial fraction of the binaries have significantly deformed primaries. \citet{Pravec2016} presented a more recent and updated survey of binary asteroids, and found that many secondaries had elongated shapes, and that asynchronous secondaries were more common in systems with more eccentric and wider companions. The triple system (87) Sylvia is composed of a primary with an irregular shape best fit by an ellipsoid with $a/b\sim1.46$ and two moons, Romulus and Remus \citep{Marchis2005}. The outer moon, Romulus is highly elongated with $a/b\sim2.7$ \citep{Berthier2014}. The main belt asteroid (216) Kleopatra is perhaps the best example of a system with an irregularly shaped primary. Radar observations of Kleopatra presented in \citet{Ostro2000} revealed that it had an extreme dumbbell shape. \citet{Descamps2011} presented Keck and Spitzer observations of the  system that identified two small Satellites, now named Cleoselene and Alexhelios. Their data also confirmed the extreme ``dog-bone" shape of the primary of (216) Kleopatra, which was refined by \citet{Shepard2018}. The aspect ratio of the interstellar object 1I/`Oumuamua was estimated to to be $6:1$ by \citet{Jewitt2017}, and even as large as  $10:1$ \citep{Meech2017} based on its brightness fluctuations. The shape of `Oumuamua was later demonstrated to be a  $115 \times 111 \times 19$ m ellipsoid by \citet{Mashchenko2019}. 

While these  studies have revealed the ubiquity, shape and formation mechanisms of deformed binaries, characterizing the mutual gravitational interaction of these bodies is challenging. Dynamical models of the rotational states of each body is non-trivial, because the permanent asphericity of both bodies strongly perturbs their orbital dynamics \citep{Scheeres2002}. \citet{Cuk2010} found that satellites in small binaries were especially susceptible to chaotic rotation, even when considering low eccentricity orbits and effects of YORP.  \citet{Nadoushan2015} considered the rotational dynamics of binary systems in which the satellite was permanently deformed, and quantified how  a satellite was  susceptible to chaotic dynamics via the overlap of the spin-orbit resonances. They showed  that as the satellite's asphericity and eccentricity increased, it was  susceptible to strong chaos (where ``chaotic regions are large compared to the non-chaotic regions", as defined in the Conclusion section of  \citet{Nadoushan2015}).  In two  studies, \citet{Nadoushan2016} and \citet{Batygin2015} used a Hamiltonian approach to demonstrate that if not only the secondary, but also the primary body can sustain a significant quadruple moment, then in addition to spin-orbit resonances, the phase space portrait of the system is augmented with spin-spin resonances. This yields an even more sophisticated evolutionary picture. The analysis of \citet{Batygin2015}, however, is largely analytic, and is restrictive in scope by its exploratory design. \citet{Nadoushan2016} explored spin-orbit and spin-spin resonances in the context of the ellipsoidal full two body problem, while \citet{Hou2017} included the variations in the  mutual orbit between the two bodies which may be appropriate for similarly sized and close minor planets. Haumea has one of the largest quadrupole moments for bodies of its size and a fast $\sim3.9$ hour spin period, which makes its satellites seemingly strong candidates for exotic the spin-spin resonances investigated in \citet{Batygin2015}. However, \citet{Hastings2016} presented Hubble and Magellan observations that demonstrated that Hi'iaka, the largest Satellite of Haumea, had a $\sim9.8$ hour spin period, about $\sim120$ times faster than its orbital period, and was likely not in one of these spin-spin resonances. \citet{Lages2017} used a dumbbell model to approximate the quadrupole moment of Ida, and demonstrated that the satellite Dactyl was very close to  the chaotic region. \citet{Correia2018} demonstrated that the plutino triple system (47171) Lempo was chaotic on a very short timescale by modeling both the orbital and spin dynamics. \citet{Naidu2015} performed numerical simulations of the spin dynamics of several well characterized near-Earth asteroid binaries and found that asynchronous rotators had large regions of chaotic phase space.

Upcoming observatories and missions offer a promising route to characterizing the intricate spin dynamics of these small deformed binaries. While systems where both the primary and satellite have  dramatic deformations are uncommon in the Solar System, the existence of contact binaries such as (216) Kleopatra and 2014 MU69 \citep{Stern2019} imply that future surveys will discover similarly deformed objects.  \citet{Durech2005} presented synthetic photometric observations of binary objects with rotation periods ranging from 2-24 hours. Many of their observational assumptions are well-suited to the upcoming Vera Rubin Observatory, and they found that for the cases of ``fast and slow rotating asteroids, tumbling asteroids, and binary objects", ``none of these cases can be misinterpreted as a principal axis rotator." These results imply  that the Vera Rubin Observatory will be able to identify the existence of chaotic rotational dynamics in binary objects. The \textit{Janus} \citep{Scheeres2020} mission was recently selected as a NASA SIMPLEx mission, and offers the opportunity for  unprecedented \textit{in-situ} observations of  binary objects. \textit{Janus} will visit two binary near-Earth asteroids, 1991 VH, which is in a chaotic  spin state, and 1996 FG3, which is in a singly synchronous stable configuration. Therefore, understanding the intricate chaotic dynamics related to spin-orbit and spin-spin coupling states from a theoretical standpoint is imperative.  In recent studies, simplified models of  gravitationally interacting bodies, such as those using rigid rods \citep{Kwiecinski2018} or concentric point masses \citep{Batygin2015}, have illuminated intricate dynamics. In this work we continue the line of investigation in \citet{Batygin2015} and relax some of the simplifying assumptions employed in previous studies. In particular, we characterize the emergence of chaos due to both spin-orbit, as well as spin-spin resonances, and derive the relevant criteria, which will be useful for constraining the properties of chaotic binaries. As a case study, we apply the criteria to (216) Kleopatra, and find that the two satellites, Cleoselene and Alexhelios, could quite plausibly satisfy the criteria for chaotic rotation due to the overlap of both the spin-orbit resonances and the retrograde spin-spin resonances.

This paper is organized as follows. In \S\ref{Section:torque}, we describe the analytic ``two-dumbbell" model,  calculate the torque on the satellite, and derive a Hamiltonian governing the spin dynamics. In \S\ref{section:prograde}, \S\ref{section:spinorbit}, and \S\ref{section:retrograde}, we investigate the prograde spin-spin, spin-orbit and retrograde spin-spin resonances individually, and derive the criteria for chaotic rotation for each. In \S\ref{section:overlap}, we examine how all of these individual resonances interact with each other. In \S\ref{section:kleopatra}, we apply our criteria to (216) Kleopatra and conclude. All of the code that generated the figures in this paper and a mathematica notebook containing the derivations of the relevant equations may be found at \url{https://github.com/DSeligman/Onset_Chaos_SpinSpinOrbit}.


\section{Evaluation of the Torque }\label{Section:torque}

\begin{figure}
\begin{center}
\includegraphics[scale=0.7,angle=0]{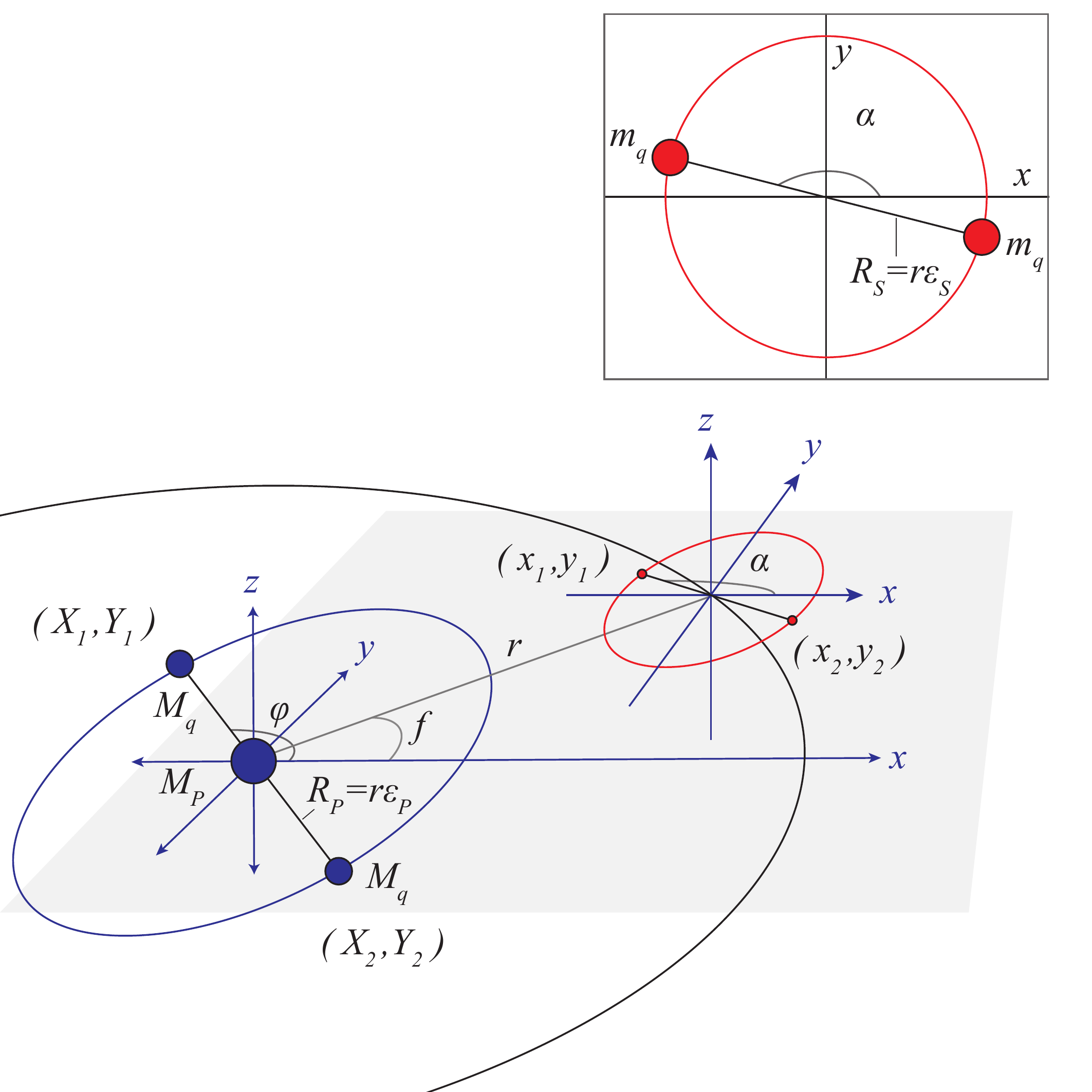} 
\caption{A schematic diagram showing the geometric setup of the proposed analytic model. The quadrupole moment arising from the permanent aspheroidal shape of both the primary and satellite is approximated by two diametrically opposed point masses. The primary is drawn in purple, and the satellite is drawn in red. The colored circle connecting the point masses represents  the effective quadrupole moments averaged over full circulations of the rotation angles, $\phi$ and $\alpha$, for the primary and satellite, respectively. The remainder of the angles and parameters are defined in the text.}\label{Fig:2dSchematic}
\end{center}
\end{figure}

We consider an idealized model for a gravitationally interacting primary and satellite pair. In Figure \ref{Fig:2dSchematic}, we show a schematic diagram for the geometric setup of the problem. Our model is simplified by construction, in order to be easily applicable to any pair of objects that are able to sustain  permanent aspheroidal shapes. In order to capture the intricate rotational dynamics that arise from the tidal gravitational forces, we approximate the  quadrupole moments of both bodies by two diametrically opposed point masses, equidistant from the center of each body. In order to capture the spin-orbit dynamics,  we also include a central point mass for the primary, which we neglect for the satellite.

The parameters of the model are defined as follows. The primary and satellite  center of masses are separated by an orbital distance, $r$. The satellite's orbit is denoted in the figure, and has eccentricity, $e$, and  semi-major axis, $\bar{r}$. The position of the satellite along its orbit is given by the true anomaly, $f$. The unprimed, $(x,y)$, coordinates indicate a stationary reference frame. The quadrupole moment of the primary is displayed in purple, and the positions of two point masses with mass $M_q$ are located at coordinates $(X_1,Y_1)$ and $(X_2,Y_2)$, denoted by capital letters. The radius of the primary, which is the distance between the origin and each mass, is given by $R_P$. The ratio of the size of the primary to the instantaneous orbital separation is denoted by $\epsilon_P=R_P/r$. It is important to note that this ratio depends on the true anomaly and is time dependant.  In this formulation, the two point masses do not account for the entire mass of the primary, which we denote as $M_P$ and  include in order to capture the relative importance of the spin-orbit dynamics to the spin-spin resonances of the satellite. The level of deformation of the primary can be related to the relative strength of the quadrupole moment, given by the proxy $M_q$, with respect to $M_P$. For many mildly elongated primary objects, $M_q<<M_P$, while $M_q$ can be a substantial fraction of $M_P$ for a body like Kleopatra. Using the three moments of inertia $A,B,C$, for the dumbbell model $B_P-A_P=2M_q R_P^2$ \citep{Murray1999}, and $C_P=2M_PR_P^2$. Here we have followed the notation where $A<B<C$, as was presented in Section 7 of \citet{Naidu2015}, and these are analogous to the moments of inertia for the satellite used in \citet{Nadoushan2015}, $I_1$, $I_2$ and $I_3$, where $I_1<I_2<I_3$. The instantaneous orientation of the primary is denoted by the angle, $\phi$. The purple circle that intersects the two point masses denotes the effect of the  quadrupole moment of the body averaged over full circulations of its rotation, $\phi$.



The satellite has variables that are defined similarly. It is also constructed with two concentric point masses, with mass $m_q$, which are located equidistant from the position along the orbit, $(x_0,y_0)$. The position of the two masses are located at coordinates $(x_1,y_1)$ and $(x_2,y_2)$,  denoted by lower case letters. The radius is $R_S$, and the ratio of the size of the satellite to the orbital separation is  $\epsilon_S=R_S/r$. The angle $\alpha$ is the angle from the unprimed $x-$ axis to the satellite orientation, illustrated in the zoom-in in the upper right hand corner of Figure \ref{Fig:2dSchematic}. The mass of the satellite is given by $M_S$, and $m_q=(B_S-A_S)/(2R_S^2)$. We have not included the central point mass of the satellite in this analysis, because the torque exerted on this central mass is small, and therefore, it does not influence the spin-orbit or spin-spin dynamics of the satellite.

The positions of the four point masses in the unprimed frame have numbered subscripts. The positions of the satellite's point masses are,
\begin{equation}
    x_1=r\cos (f)+\epsilon_sr\cos(\alpha)\, \,, \, \, y_1 = r\sin (f)+\epsilon_sr\sin(\alpha)\,,
\end{equation}
\begin{equation}\label{eq:x2y2}
    x_2=r\cos (f)+\epsilon_sr\cos(\alpha+\pi)\, \,, \, \, y_2 = r\sin (f)+\epsilon_sr\sin(\alpha+\pi)\,.
\end{equation}
Similarly, the position of the primary's masses are given by,
\begin{equation}
    X_1=\epsilon_pr\cos(\phi)\, \,, \, \, Y_1 = \epsilon_pr\sin(\phi)\,,
\end{equation}
\begin{equation}
    X_2=\epsilon_pr\cos(\phi+\pi)\, \,, \, \, Y_2 = \epsilon_pr\sin(\phi+\pi)\,,
\end{equation}
The gravitational potential between any two masses will be of the form,
\begin{equation}
    \mathcal{U}=-\frac{G M}{(x^2+y^2)^{1/2}} \,.
\end{equation}
We calculate the acceleration on both satellite masses, where the first subscript number indicates which satellite mass the acceleration is calculated at, and the second subscript number indicates which primary mass is contributing to the force. We also include the acceleration on each satellite mass from the central mass which we denote with a $0$ subscript. These six accelerations are given by,
\begin{equation}\label{Equation:GradU10}
   \nabla \mathcal{U}_{10}=-\frac{G M_P}{(x_1^2+y_1^2)^{3/2}}(x_1\, \,, \, \, y_1) \,,
\end{equation}

\begin{equation}\label{Equation:GradU11}
   \nabla \mathcal{U}_{11}=-\frac{G M_P}{((x_1-X_1)^2+(y_1-Y_1)^2)^{3/2}}(x_1\, \,, \, \, y_1) \,,
\end{equation}
\begin{equation}
   \nabla \mathcal{U}_{12}=-\frac{G M_q}{((x_1-X_2)^2+(y_1-Y_2)^2)^{3/2}}(x_1\, \,, \, \, y_1) \,,
\end{equation}

\begin{equation}
   \nabla \mathcal{U}_{20}=-\frac{G M_q}{(x_2^2+y_2^2)^{3/2}}(x_2\, \,, \, \, y_2) \,,
\end{equation}

\begin{equation}
   \nabla \mathcal{U}_{21}=-\frac{G M_q}{((x_2-X_1)^2+(y_2-Y_1)^2)^{3/2}}(x_2\, \,, \, \, y_2) \,,
\end{equation}
and
\begin{equation}\label{Equation:GradU22}
   \nabla \mathcal{U}_{22}=-\frac{GM_q}{((x_2-X_2)^2+(y_2-Y_2)^2)^{3/2}}(x_2\, \,, \, \, y_2) \,.
\end{equation}

We calculate the rotational torque, $\tau$, on the satellite from each acceleration defined in Equations \ref{Equation:GradU10}-\ref{Equation:GradU22}.  We neglect the inertial terms that affect the orbital problem but not the spin-orbit problem, and our resulting Hamiltonian  reduces to the standard spin-orbit problem.  The resulting torque from each primary-satellite point mass pair, $\tau_{ij}$, is given by $\tau_{ij}=\tilde{r}_j \times m_q\nabla U_{ij} $, where $\tilde{r}_j$ is the vector pointing from the center of the satellite to the mass, and the subscripts are defined as in Equations \ref{Equation:GradU10}-\ref{Equation:GradU22}. For the mass $m_1$, $\tilde{r}_1=(r\epsilon_s \cos( \alpha), r\epsilon_s \sin (\alpha))$ and for the second mass $\tilde{r}_j=(r\epsilon_s \cos( \alpha+\pi), r\epsilon_s \sin (\alpha+\pi))$. The total torque on the satellite results from summing all  six resulting torques. After summing these terms, and truncating the expression to second order in $\epsilon_P$ and $\epsilon_S$, the resulting equation is

\begin{equation}\label{Equation:Torque}
    \tau = \frac{3G m_q\epsilon_S^2}{4r}\bigg( 2(2M_P+(4+5\epsilon_P^2)M_q)\sin(2f-2\alpha)+\epsilon_P^2M_q\big(-3\sin(2\alpha-2\phi)+35\sin(4f-2(\alpha+\phi))\big)\bigg)\, .
\end{equation}
We integrate the torque over all rotation of the satellite, $\alpha$, to get an equation for the rotational energy stored in the system.
  The integrated torque yields a straightforward Hamiltonian for the system,
\begin{equation}\label{Equation:Hamiltonian_original}
    \mathcal{H}=\frac{\mathcal{A}^2}{2}-\frac{3G }{8r^3}\bigg(\frac{B_S-A_S}{C_S}\bigg)\bigg( (2M_P+(4+5\epsilon_P^2)M_q)\cos(2f-2\alpha)+\frac12\epsilon_P^2M_q\big(3\cos(2\alpha-2\phi)+35\cos(4f-2(\alpha+\phi))\big)\bigg)\, , 
\end{equation}
where $\mathcal{A}$ is the spin of the satellite, $\mathcal{A}=\dot{\alpha}$ and is  conjugate to $\alpha$, and we have substituted the definition for $\epsilon_S$ into Equation \ref{Equation:Hamiltonian_original}. This Hamiltonian will govern the rotational dynamics investigated in the next several sections. 

Each term in Equation \ref{Equation:Hamiltonian_original}  governs a distinct physical effect. It is worth noting that this analysis is similar to the evaluation of the torque in \citet{Batygin2015}, with the important difference being that we employ the concentric mass model for the primary instead of using MacCullagh's formula. Comparing the three terms in Equation \ref{Equation:Hamiltonian_original}  with Equation 7 in \citet{Batygin2015} shows that the two approaches result in similar harmonics, perhaps with divergence in coefficients at higher order of $\epsilon_P$ and $\epsilon_S$.  The first term is $\sim\cos(2f-2\alpha)$, and governs the well studied spin-orbit coupling, since it resonates when the spin of the satellite is commensurate with the orbit. This has a coefficient which depends only on the mass of the satellite, and a term that depends on the quadrupole moment of the primary, and the entire coefficient scales with the asphericity of the satellite. In the limit where the satellite has no quadrupole moment, the spin-orbit coupling disappears. 

The other two terms govern the more subtle spin-spin resonances, which depend on a higher order of the size of the primary. The second term is $\sim \cos(2\alpha-2\phi)$, and librates when the rotation rate of the secondary is commensurate with the rotation rate of the primary ($\dot{\phi}\cong\dot{\alpha}$). When the rotation rates are decoupled from the orbit $\dot{f}<<\dot{\phi}$, then the third term librates when the two bodies have opposite spin ($\dot{\phi}\cong-\dot{\alpha}$). The second term governs the prograde  spin-spin resonance, while the third term governs the retrograde spin-spin resonance.  In the limit where the primary is not at all deformed, where $\epsilon_P\rightarrow0$ and $M_q\rightarrow0$, we see that only the spin-orbit resonance remains. This indicates that in the case where the primary is not deformed and only the satellite is deformed, only the spin-orbit resonance exists. As the primary becomes more deformed and is able to sustain  a permanent quadrupole moment, the spin-spin prograde and retrograde resonances appear, and the strength of the spin-orbit resonance increases as $M_q\rightarrow M_P$. However, if the satellite has no quadrupole moment and $B_S-A_S\rightarrow0$, then none of the  spin-orbit or spin-spin resonances exist.

In the next three sections, we will investigate the dynamics arising from each of these terms individually. In order to do this, we consider cases where  each of the resonances described in the three terms in Equation \ref{Equation:Hamiltonian_original} are well separated in frequency space. If the satellite is in one of the resonances governed by these terms, then  we may make the approximation that the other linear combinations of angles in Equation \ref{Equation:Hamiltonian_original} vary from $0$ to $2\pi$ very quickly compared with the resonant term, and average to zero over many rotation cycles of the resonance. Because we employ this averaging method, we can investigate the rotational dynamics of each term in isolation. Although we defer our discussion of the system (216) Kleopatra until \S \ref{section:kleopatra}, we note that these three resonances are well separated in this instance, as can be seen in Figure \ref{Fig:PhasespaceKleo}.

\section{Prograde Spin-Spin Resonance}\label{section:prograde}

In this section, we only consider the dynamics influenced by the term in the Hamiltonian, Equation \ref{Equation:Hamiltonian_original}, that governs the 1:1 prograde spin-spin resonance. This approach is justified by employing the averaging technique described in the final paragraph of the last section. We consider the reduced Hamiltonian, $\mathcal{H}_{pro}$, which only contains the term proportional to $\sim\cos(2\alpha-2\phi) $,

\begin{equation}\label{Equation:Hamiltonian_prograde}
    \mathcal{H}_{pro}=\frac{\mathcal{A}^2}{2}-\frac{9GM_q \epsilon_P^2}{16r^3}\bigg(\frac{B_S-A_S}{C_S}\bigg)\cos(2\alpha-2\phi)\, . 
\end{equation}

The plan of this section is as follows. In Subsection \ref{subsection:Circular}, we  investigate the dynamics of this Hamiltonian for the case of a purely circular orbit with zero eccentricity. We  show via a contact transformation that this Hamiltonian reduces to the equations of a pendulum, with a nonzero offset centered at the rotation rate of the primary. We show the contact transformation in detail to quantify the location of the separatrix in phase space. In future sections, we  employ similar contact transformations and omit many of the intermediate steps for brevity.   In Subsection \ref{subsection:spinspineccentric}, we  introduce the eccentricity of the orbit, and show that it splits the resonance into three resonances, rendering the system similar to a modulated pendulum. 

\subsection{Spin-Spin Resonance for a Circular Orbit}\label{subsection:Circular}

In this subsection, we  consider the spin evolution of a satellite on a  circular orbit, where the separation between the primary and satellite is constant in time, $r(t)=r$. We define the rotation of the primary to be $\Omega_p$, where $\phi=\Omega_p t$,  such that the primary is undergoing steady state rotation. This assumption reduces the problem to a one-degree-of-freedom system, because the primary does not feel the back reaction of the torques from the satellite. This approximation reduces the complexity of the problem and is valid when the mass of the primary is much larger than the mass of the satellite, $M_P>>M_S$. We transform the variable for time, $t$, to  an angle, $t=\theta$,  that scales linearly with time. By extending phase space to include the conjugate momentum to this new angle, $\Theta$, Equation \ref{Equation:Hamiltonian_prograde} becomes,
\begin{equation}\label{Eq:Hamiltonian_Prograde_extended}
    \mathcal{H}_{pro}=\frac{\mathcal{A}^2}{2}-\frac{9GM_q \epsilon_P^2}{16r^3}\bigg(\frac{B_S-A_S}{C_S}\bigg)\cos(2\alpha-2\Omega_p\theta)+\Theta\, .
\end{equation}
We now utilize a type-2 contact transformation to define the generating function $\mathcal{F}_2$,

\begin{equation}
    \mathcal{F}_2=(\alpha-\Omega_p\theta)\Gamma+\theta \Xi\, .
\end{equation}
This contact transformation defines a mapping from the two sets of canonical actions and angles, where  $\mathcal{F}_2:(\alpha,\mathcal{A}),(\theta,\Theta)\xrightarrow{}(\gamma,\Gamma),(\xi,\Xi)$. The new angles are defined as $\xi=\theta$ and $\gamma = \alpha-\Omega_p\theta$, and the new actions correspond to these new angles. The new actions follow from the transformation equations:

\begin{equation}
    \mathcal{A}=\frac{\partial\mathcal{F}_2 }{\partial \alpha}=\Gamma\, ,
\end{equation}
and
\begin{equation}
    \Theta=\frac{\partial\mathcal{F}_2 }{\partial \theta}=\Xi -\Omega_p\Gamma\, .
\end{equation}
We substitute these into our reduced Hamiltonian, Equation \ref{Eq:Hamiltonian_Prograde_extended}, which gives,
\begin{equation}
    \mathcal{H}_{pro}=\frac{\Gamma^2}{2}-\frac{9GM_q \epsilon_P^2}{16r^3}\bigg(\frac{B_S-A_S}{C_S}\bigg)\cos(2\gamma)+\Xi -\Omega_p\Gamma\, . 
\end{equation}
Here, we note that Hamilton's equations indicates that $\frac{d \Xi}{dt}=-\frac{\partial \mathcal{H}}{\partial \xi}=0$, so $\Xi$ is a constant in time. Therefore, we can subtract this constant and, 
\begin{equation}
    \mathcal{H}_{pro}=\frac{\Gamma^2}{2}-\frac{9GM_q \epsilon_P^2}{16r^3}\bigg(\frac{B_S-A_S}{C_S}\bigg)\cos(2\gamma) -\Omega_p\Gamma\, . 
\end{equation}
By completing the square, we  add the constant $\Omega_p^2/2$ to the Hamiltonian , yielding
\begin{equation}\label{Eq:Hamiltonian_prograde_transformed}
    \mathcal{H}_{pro}=\frac12(\Gamma-\Omega_p)^2-\frac{9GM_q \epsilon_P^2}{16r^3}\bigg(\frac{B_S-A_S}{C_S}\bigg)\cos(2\gamma) \, . 
\end{equation}
By performing this transformation, we have transformed into the frame that is rotating with the primary. We can examine some features of this Hamiltonian. If we apply Hamilton's equation to Equation \ref{Eq:Hamiltonian_prograde_transformed}, we see,

\begin{equation}
    \frac{d\gamma}{dt}=\frac{\partial \mathcal{H}_{pro}}{\partial\Gamma}= \Gamma-\Omega_p\, .
\end{equation}
From our definition, $\gamma=\alpha-\phi=\alpha-\Omega_P t$. This implies that $\dot{\gamma }= \dot{\alpha}-\dot{\phi}=\dot{\alpha}-\Omega_P$. Therefore $\Gamma = \frac{d\alpha}{dt}$. The fixed point of the Hamiltonian is where $\Gamma=\mathcal{A}=\Omega_P$, which implies that $\dot{\alpha}=\Omega_P$, or  $\dot{\alpha}=\dot{\phi}$. Therefore, this is the 1:1 spin-spin prograde resonance of the primary and secondary. 

\begin{figure}
\begin{center}
\includegraphics[scale=0.35,angle=0]{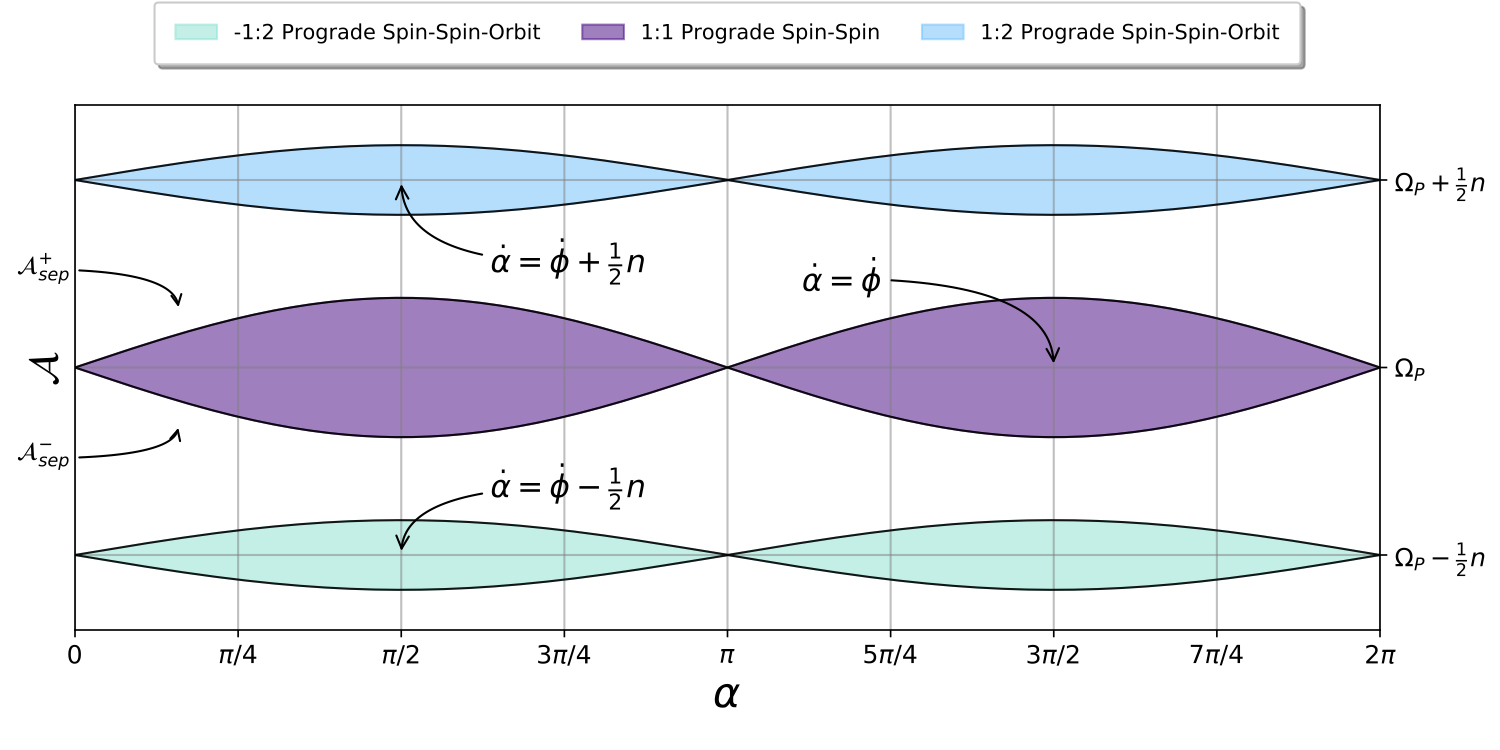} 
\caption{A schematic phase space diagram showing the location of the three resonances arising from the eccentric orbit splitting  the 1:1 prograde spin-spin resonance. The three separatrices resulting from the decomposition of each of the terms in Equation \ref{Equation:FullSpinSpin} are plotted in black lines. The purple shaded region corresponds to the regions of phase space where the satellite is in the 1:1 spin-spin resonance, characterized by $\mathcal{A}=\dot{\alpha}=\Omega_P$. The modulation of the orbital separation from the eccentric orbit acts analogously as a pendulum on a spring, splitting this resonance into three resonances, where the two new resonances are shown in turquoise and light blue and are separated by $\pm n/2$. In this figure, we assume  the primary is much more massive than the satellite, with an orbital eccentricity $e=0.1$ and primary size ratio $\epsilon_P=.3$, for which the system is  below the criteria for resonance overlap provided in Equation \ref{eq:chaos_criteria_spinspinprograde}.}\label{Figure:Schematic_Prograde}
\end{center}
\end{figure}

This  Hamiltonian is quite clearly the same as that for a pendulum, centered at $\mathcal{A}=\Omega_p$ with a resonance half-width of $2\sqrt{\frac{9GM_q R_P^2}{16\bar{r}^5}\big(\frac{B_S-A_S}{C_S}\big)}$. The separatrix of this Hamiltonian is shown in the center of the schematic phase space diagram in Figure \ref{Figure:Schematic_Prograde}. The purple shaded region corresponds to the librations around the $1:1$ spin-spin resonance in phase space. 

\begin{figure}
\begin{center}
\includegraphics[scale=0.35,angle=0]{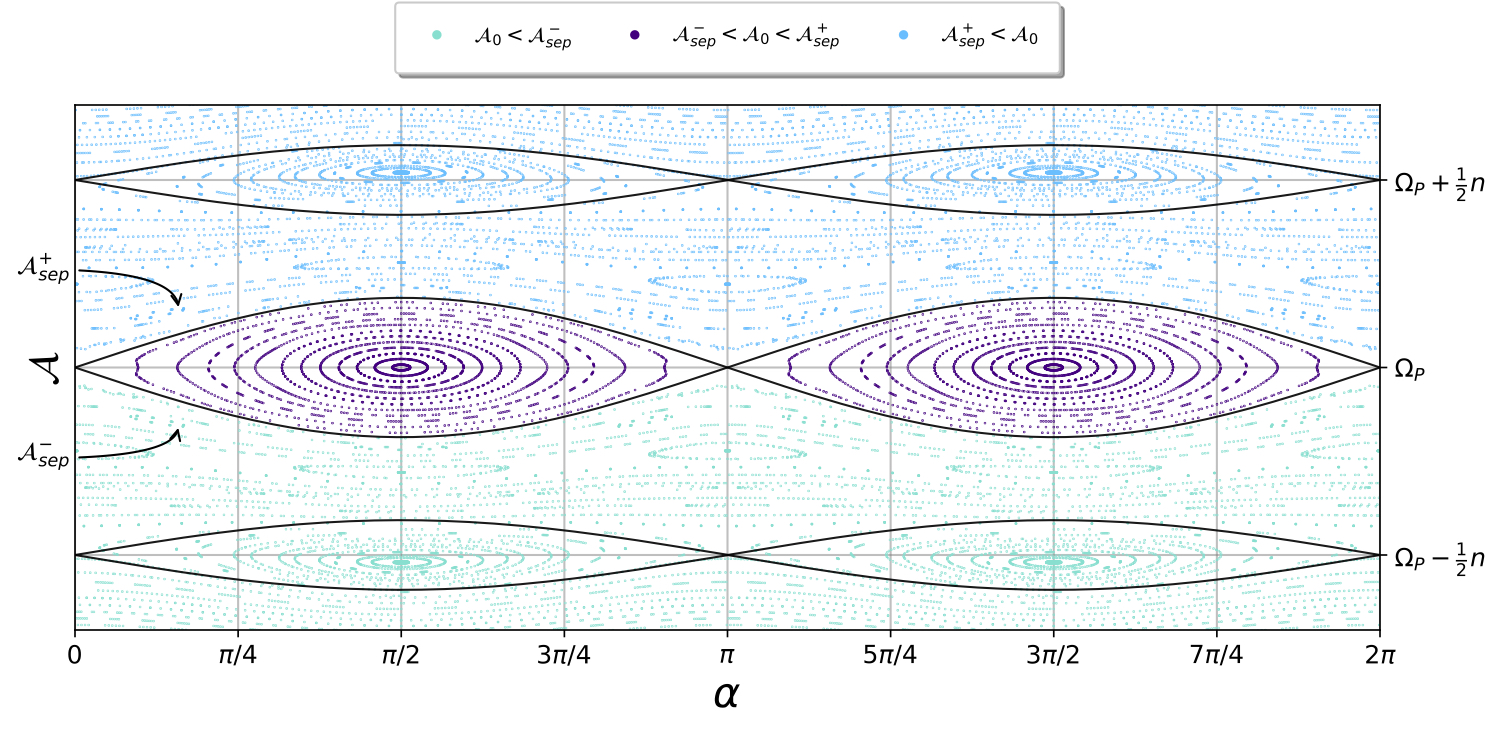} 
\caption{A surface of section for the Hamiltonian for initial trajectories centered close to the fixed point of  the eccentric  spin-spin prograde resonances. The binary has the same parameters as in Figure \ref{Figure:Schematic_Prograde}, and the same separatrices are overplotted here. The eccentricity, orbital period and separation are chosen such that the resonances are separated and not overlapping in phase space, and  the widths of the upper and lower resonance are smaller than that of the central spin-spin resonance. The color of the points indicates where the trajectory was initialized in phase space. Turquoise, purple and light blue points correspond to trajectories that were initialized below, within and above the separatrix of the 1:1 spin-spin resonance. There is no chaos in the surface of section, and smaller sub-resonances  appear due to higher order resonances in the Hamiltonian.}\label{Figure:ProgradeSurface}
\end{center}
\end{figure}

\subsection{Spin-Spin Resonances for an Eccentric Orbit}\label{subsection:spinspineccentric}

In the previous subsection, we  only considered the case of a circular orbit where the distance between the primary and satellite is constant in time. In this subsection, we show that if the satellite has an eccentric orbit, then the 1:1 spin-spin prograde resonance splits into three resonances, separated in frequency by half of the mean motion.  For the case of an eccentric orbit, $r$ now depends on time. Using a similar equation to Equation 2.83 in \citet{Murray1999}, by Taylor expanding we see that, to first order in eccentricity,

\begin{equation}\label{Equation:r5_expansion}
    \frac{1}{r^5}=\frac{1}{\bar{r}^5}(1+5e\cos M+\mathcal{O}(e^2))\, ,
\end{equation}
where $M=nt$ is the mean anomaly,  $n=2\pi/T=\sqrt{G(M_P+M_S)}/\bar{r}^{3/2}$ is the mean motion, and $T$ is the period of the orbit. Substituting our definition of $\epsilon_P=R_P/r$\footnote{Note that in our construction, $\epsilon_P$ is formally a function of time for the eccentric orbit.} and the expansion given in Equation \ref{Equation:r5_expansion}, the Hamiltonian in Equation  \ref{Equation:Hamiltonian_prograde} can then be written as,

\begin{equation}\label{eq:modulatedpend}
    \mathcal{H}_{pro}=\frac{\mathcal{A}^2}{2}-\frac{9GM_q R_P^2}{16\bar{r}^5}\bigg(\frac{B_S-A_S}{C_S}\bigg)(1+5e\cos M)\cos(2\alpha-2\phi)\, ,
\end{equation}
to first order in the eccentricity of the orbit. 

To understand the dynamics from a qualitative perspective, we remind the reader of the modulated pendulum, or a pendulum with a base suspended on a spring, as a paradigm for chaotic dynamics. In this simple example, the effective value of gravitational acceleration, $g$, is replaced by, $g(1+\cos(\omega t))$, where $\omega$ is the frequency set by the properties of the spring.  Upon multiplying out the terms, we see how a Hamiltonian with a single harmonic splits into one with three harmonics, each separated in frequency by $\pm \omega/2$. This physical analogy is useful to keep in mind, when we expand Equation \ref{eq:modulatedpend} using trigonometric substitutions to obtain:

\begin{equation}\label{Equation:FullSpinSpin}
        \mathcal{H}_{pro}=\frac{\mathcal{A}^2}{2}-\frac{9GM_q R_P^2 }{16\bar{r}^5}\bigg(\frac{B_S-A_S}{C_S}\bigg)\bigg(\cos(2\alpha-2\phi)+\frac52e\cos(2\alpha-2\phi-M)+\frac52e\cos(2\alpha-2\phi+M)\bigg)\, .     
\end{equation}

The first term in this Hamiltonian  governs the spin-spin coupling in the absence of the eccentricity modulation, and this equation reduces to the equation for a circular orbit in the case of $e=0$. The other terms, which are first order in eccentricity, represent modulations to these oscillations that are offset from the spin-spin resonance by $\frac12n$.  We note that, in isolation, each one of these resonances constitutes a pendulum-like Hamiltonian, and orbital eccentricity is nothing more than a modulation of the strength of gravitational torques felt by the satellite, with the mean motion as the driving frequency. Thus, by analogy, we can expect the harmonic to split into three, separated by $\pm n/2$. 

In the interest of brevity, we will not show the contact transformations that generate the fixed points of these split resonances. However,  contact transformations such as the one presented in Subsection \ref{subsection:Circular} can be performed to show that the two sub-resonances are centered at $\mathcal{A}=(\Omega_p\pm\frac{n}{2})$, and both have a resonance half-width of $2\sqrt{\frac{45GM_q R_P^2e}{32\bar{r}^5} \big(\frac{B_S-A_S}{C_S}\big)}$. By the merits of the contact transformation performed in Subsection \ref{subsection:Circular}, it should be straightforward to see this by examination of the harmonics and coefficients of the second and third terms in Equation \ref{Equation:FullSpinSpin}.  We have demonstrated that the eccentric orbit splits the 1:1 spin-spin resonance into three resonances, centered at synchronous rotation between the primary and satellite, $\dot{\alpha}=\Omega_P$, with rotation rates of the satellite being $\dot{\alpha}=\Omega_P-n/2$ and $\dot{\alpha}=\Omega_P+n/2$.

 Using what is colloquially referred to as the  ``Chirikov criterion," when the resonances overlap, or the widths of the resonances are greater than the distance between their centers, the system will exhibit chaotic dynamics \citep{Chirikov1979}. This criteria has been widely used to predict the onset of chaotic dynamics in the spin-orbit problem, and was used in the original \citet{Wisdom1984} study to predict the tumbling of Hyperion, and in Section 4 of \citet{Nadoushan2015}. We follow a similar precedent as presented in \citet{Nadoushan2016}, who first applied the Chirikov criterion to predict the onset of chaos for the spin-spin problem. The criteria for the onset of chaos, or for resonance overlap, will be

\begin{equation}
    \frac{n}{2}<2\sqrt{\frac{9GM_q R_P^2}{16\bar{r}^5}\bigg(\frac{B_S-A_S}{C_S}\bigg)}+2\sqrt{\frac{45GM_q R_P^2e}{32\bar{r}^5}\bigg(\frac{B_S-A_S}{C_S}\bigg) }\,.
\end{equation}
Since the mean motion is given by $n=\sqrt{G(M_P+M_S)}/\bar{r}^{3/2}$, the criteria for chaos for the spin-spin resonance can be written as

\begin{equation}\label{eq:chaos_criteria_spinspinprograde}
    \frac{\bar{r}}{R_P}<3 \sqrt{\frac{M_q}{M_P+M_S}}\sqrt{\bigg(\frac{B_S-A_S}{C_S}\bigg)}(1+\sqrt{\frac{5e}{2}})\,.
\end{equation}
While resonance overlap can occur at very low values of eccentricity, the fraction of phase space that will be chaotic will be zero as $e\rightarrow0$. To quantify this point, the criteria for the upper resonance to completely overlap with the central resonance, as a function of the eccentricity, is,

\begin{equation}
    \frac{2}{5}\bigg(\frac{n}{4} \bigg(\frac{9GM_q R_P^2}{16\bar{r}^5}\frac{B_S-A_S}{C_S}\bigg)^{-1/2}+1\bigg)^2<e\,.
\end{equation}

In Figure \ref{Figure:Schematic_Prograde}, we show a schematic of the phase space diagram, where the three colored sections correspond to the locations of these three resonances. We plot the separatrices for the three resulting resonances. For this configuration, we have assumed that the mass of the  primary is much greater than that of the satellite. We have also assumed that the satellite has an orbital eccentricity of  $e=0.1$ and $R_P=0.3\bar{r}$. Finally, we have assumed that the primary has a strong quadrupole moment where $M_q=M_P/5$, and that the satellite has an asphericity in its moment of inertia tensor similar to an object with an aspect ratio of $a_s/b_s\sim3.5$. We have chosen these initial conditions to make a reasonable representation of two elongated objects on a relatively close orbit, but one  such that the resonances are separated in phase space and not overlapping.  The widths of the upper and lower resonances are considerably smaller than to the width of the center resonance. Based on the criteria for resonance overlap given by Equation \ref{eq:chaos_criteria_spinspinprograde}, these values are  below what would be required for resonance overlap, as evident by the location of the separatrices. We note, however, that for other configurations of separations, orbital period and eccentricities, the resonance widths will not necessarily be comparable, and the resonances can overlap. 

In Figure \ref{Figure:ProgradeSurface}, we show a stroposcopic surface of section taken at $\Omega_Pt=2\pi,4\pi...$ for the  Hamiltonian representing the full evolution of the system.  We have plotted this with respect to the original action and angles, $\alpha$ and $\mathcal{A}$, which correspond physically to the rotational state of the satellite. The color coding of the trajectories indicate if they were initialized below, inside of or above the separatrix for the 1:1 spin spin resonance. We have chosen parameters such that the three resonances are moderately separated in phase space and such that the width of the middle resonance is stronger than the other two. We can see that  the resonances are formally separated and there are no chaotic regions in the surface of section. Due to interactions among split spin-spin harmonics, higher order resonances arise, as is made evident by the existence of  resonant islands outside of the separatrices. 


\section{Spin-Orbit Coupling}\label{section:spinorbit}

In this section, we investigate the spin dynamics of the satellite under the action of the spin-orbit coupling term in the original Hamiltonian, Equation \ref{Equation:Hamiltonian_original}.  It is important to note that the spin-orbit coupling was extensively studied in \citet{Nadoushan2015}, and our results in this section are in good agreement with the previous work. Employing the same averaging method as in the previous section, we only consider the reduced spin-orbit Hamiltonian, 

\begin{equation}
    \mathcal{H}_{orb}=\frac{\mathcal{A}^2}{2}-\frac{3G }{8r^3}\bigg(\frac{B_S-A_S}{C_S}\bigg)\bigg( (2M_P+(4+5\epsilon_P^2)M_q)\cos(2f-2\alpha)\bigg)\, , 
\end{equation}
Since the true anomaly $f$, is not a linear function of time, it is  less straightforward to perform the same analysis than for the 1:1 spin-spin coupling term. We utilize the relationship between the true anomaly and mean anomaly to rewrite the Hamiltonian as a function of mean anomaly, which is a linear function of time. Expansions for the true anomaly given by Equations 2.84 and 2.85 of \citet{Murray1999} are reproduced here:

\begin{equation}\label{Equation:cosf}
    \cos f = \cos M+e(\cos 2M-1) + \mathcal{O}(e^2)\, ,
\end{equation}
and 
\begin{equation}\label{Equation:sinf}
    \sin f = \sin M+e\sin 2M + \mathcal{O}(e^2)\, .
\end{equation}
By using trigonometric expansion and substituting these terms for the true anomaly, we recover the Hamiltonian for the spin-orbit coupling term to first order in eccentricity, 
\begin{equation}
    \mathcal{H}_{orb}=\frac{\mathcal{A}^2}{2}-\frac{3G }{8r^3}\bigg(\frac{B_S-A_S}{C_S}\bigg)\bigg(\big(2M_P+ (4+5\epsilon_p^2)M_q\big)\big(\cos (2 M-2 \alpha )- 2e \cos (M-2
   \alpha )+2e \cos (3 M-2 \alpha )\big)\bigg)\, . 
\end{equation}
The taylor expansion in eccentricity for $\frac{1}{r^3}$ is given by Equation 2.83 in \citet{Murray1999},
\begin{equation}
    \frac{1}{r^3}=\frac{1}{\bar{r}^3}(1+3e\cos M+\mathcal{O}(e^2))\, .
\end{equation}
Using this and Equation \ref{Equation:r5_expansion}, and truncating at second order in eccentricity, the Hamiltonian can be written as,
\begin{equation}\label{Equation:SpinOrbitHamiltonian}
\begin{split}
        \mathcal{H}_{orb}=\frac{\mathcal{A}^2}{2}-\frac{3G }{8 \bar{r}^3}\bigg(\frac{B_S-A_S}{C_S}\bigg)\bigg(\big(2M_P+ (4+5\epsilon_p^2)M_q\big)\cos (2 M-2 \alpha )\\- \frac{e}{2}\big(2M_P+ (4-5\epsilon_p^2)M_q\big)\cos (M-2
   \alpha )+\frac{e}{2}\big(14M_P+ (28+45\epsilon_p^2)M_q\big)\cos (3 M-2 \alpha )\bigg)\, . 
\end{split}
\end{equation}
 This equation may be directly compared to Equation 11 in \citet{Nadoushan2015}, which includes the second order terms in eccentricity, and our model  reproduces the analogous harmonics for the zeroth and first order resonances in eccentricity.

 \begin{figure}
\begin{center}
\includegraphics[scale=0.35 ,angle=0]{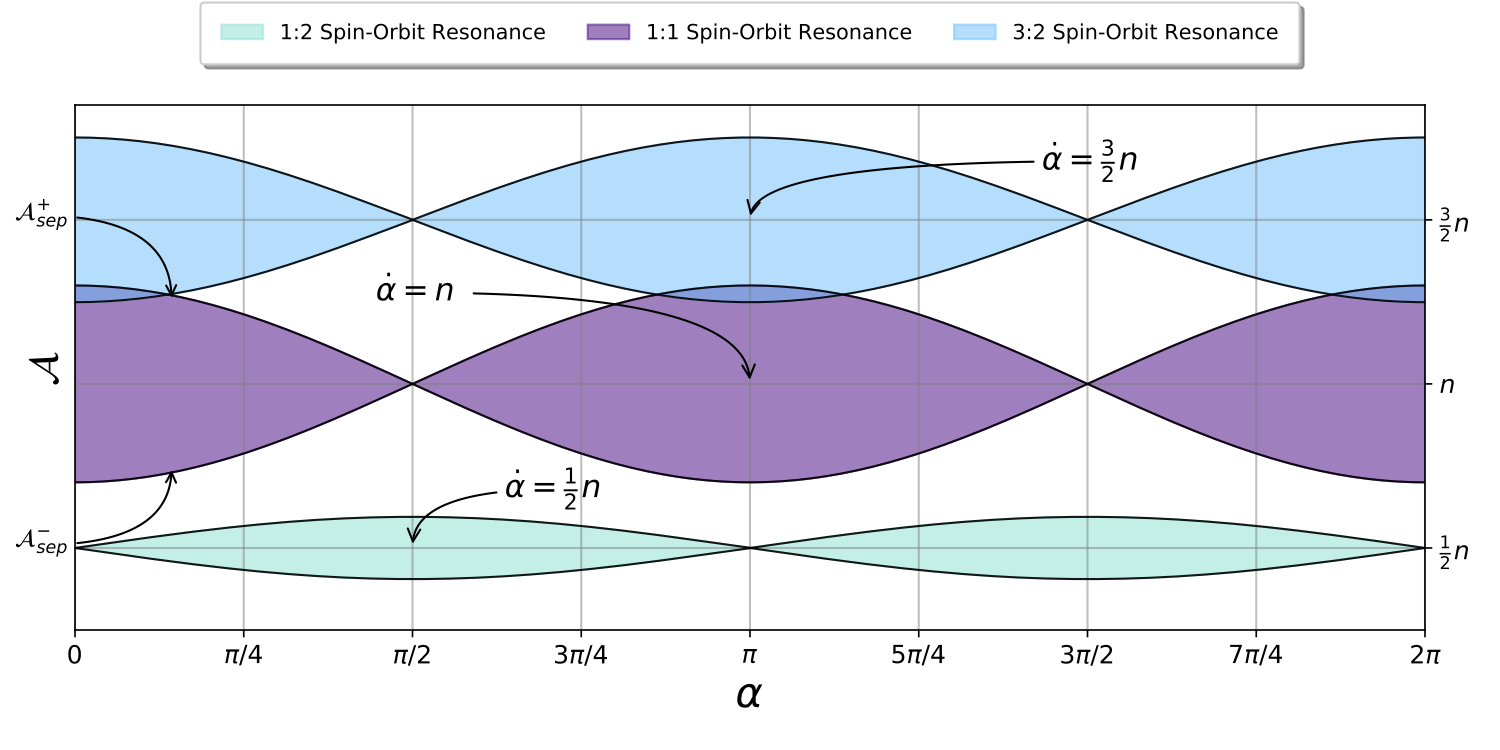}
\includegraphics[scale=0.35 ,angle=0]{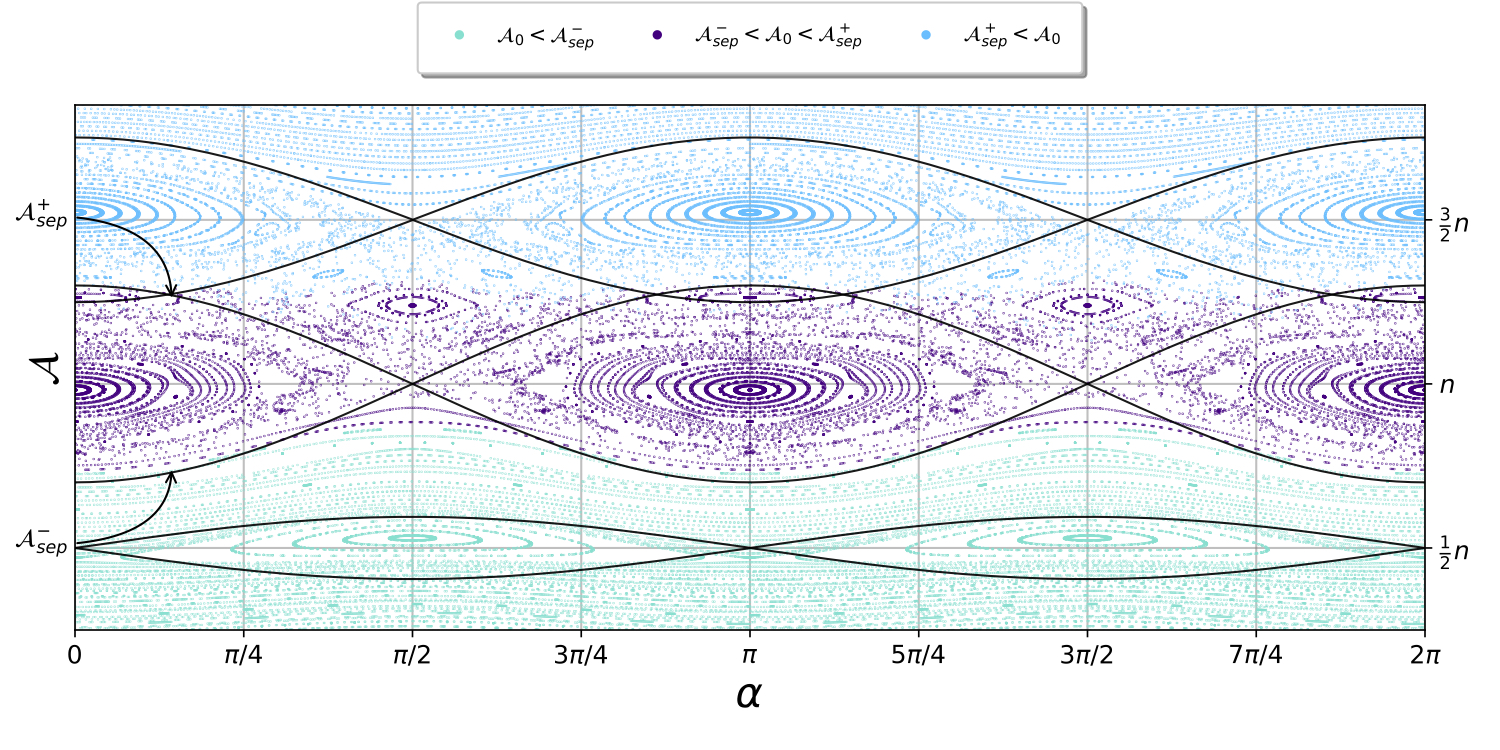}
\caption{The evolution of the spin of the satellite close to the spin-orbit resonances, given by Equation \ref{Equation:SpinOrbitHamiltonian}. The upper panel shows a schematic phase space diagram, denoting the separatrices of the three Hamiltonians governing the 1:1, 1:2 and 3:2 spin-orbit resonances derived in this section. The lower panel shows a surface of section computed for Equation \ref{Equation:SpinOrbitHamiltonian}, with the three separatrices plotted in black lines. The trajectories are color coded similarly to Figure \ref{Figure:ProgradeSurface}. We have chosen  parameters such that $M_q=M_P/2$, $e=0.2$, $R_P=0.01\bar{r}$, and $(B_S-A_S)/C_S=0.015$. }\label{Figure:Surface_spinorbit}
\end{center}
\end{figure}

With this Hamiltonian, contact transformations similar to the one used for the circular spin-spin coupling case can be employed to find the equilibrium points corresponding to  the resulting three resonances. Some of the details of the contact transformations are subtly different, but the resulting dynamics are essentially similar, so we omit the details of the transformations. The first term in Equation \ref{Equation:SpinOrbitHamiltonian} corresponds to the 1:1 spin-orbit resonance, with a fixed point where $\dot{\alpha}=n$. The half-width of this resonance is $2\sqrt{\frac{3G }{8 \bar{r}^3}\big(\frac{B_S-A_S}{C_S}\big)\big(2M_P+ (4+5\epsilon_p^2)M_q\big)}$. The eccentric orbit governs the resonances splitting in the next two terms. The second term is centered at  $\dot{\alpha}=n/2$ and has a resonance half-width of $2\sqrt{\frac{3G }{8 \bar{r}^3}\big(\frac{B_S-A_S}{C_S}\big)\frac{e}{2}\big(2M_P+ (4-5\epsilon_p^2)M_q\big)}$. Therefore, this is a 1:2 spin-orbit resonance of the primary and secondary. The third term is centered at  $\dot{\alpha}=\frac32n$, which is the 3:2 spin-orbit resonance. The half-width of this resonance is $2\sqrt{\frac{3G }{8 \bar{r}^3}\big(\frac{B_S-A_S}{C_S}\big)\frac{e}{2}\big(14M_P+ (28+45\epsilon_p^2)M_q\big)}$.  As noted in  Section 4 of \citet{Nadoushan2015}, chaotic behavior will begin when the 3:2 and 1:1 resonances begin to overlap, since they have the largest widths, and when,

\begin{equation}\label{eq:chaos_criteria_spinorbit}
    \frac{n}{2}<2\sqrt{\frac{3G }{8 \bar{r}^3}\bigg(\frac{B_S-A_S}{C_S}\bigg)}\bigg(\sqrt{\big(2M_P+ (4+5\epsilon_p^2)M_q\big)}+\sqrt{\frac{e}{2}\big(14M_P+ (28+45\epsilon_p^2)M_q\big)}\bigg)\,.
\end{equation}
As in the previous section, the fraction of phase space that will be chaotic will be zero as $e\rightarrow0$. Of particular interest, \citet{Nadoushan2015} noted that chaotic regions still appear for very small, nonzero eccentricities, but large asphericities. The criteria for the upper resonance to completely overlap with the central resonance is,

\begin{equation}
    \frac{2}{\big(14M_P+ (28+45\epsilon_p^2)M_q\big)}\bigg(\frac{n}{4} \bigg(\frac{3G }{8 \bar{r}^3}\frac{B_S-A_S}{C_S}\bigg)^{-1/2}+\big(2M_P+ (4+5\epsilon_p^2)M_q\big)\bigg)^2<e\,.
\end{equation}

The strength of the spin-orbit resonance is dominated by the mass of the primary. Only in systems where the quadrupole mass of the primary is comparable to the mass of the primary, $M_q\sim M_P/2$, will the spin-spin resonances start to overlap when the spin-orbit terms overlap. In general, the spin-orbit resonances will dominate the phase space. This implies that the intricate rotational dynamics characterized by the spin-spin resonances will only apply to aspherical Solar System binaries, such as small moons of extremely elongated objects or contact binaries, while spin-orbit coupling is a more broadly exhibited effect.

The upper panel of Figure \ref{Figure:Surface_spinorbit} shows a schematic of phase space similar to that shown in Figure \ref{Figure:Schematic_Prograde}, with the separatrices of the three relevant resonances plotted as black lines. The indigo, light blue and turquoise regions corresponds to the 1:1, 3:2 and 1:2 spin-orbit resonances, respectively.  The width of the 3:2 and 1:2 resonances are smaller than the width of the 1:1 resonance by a factor proportional to $\sqrt{e}$, because they are first order in eccentricity.  As noted by \citet{Nadoushan2015}, for any system, the width of the 1:2 resonance is   smaller than the width of the 3:2 and 1:1 resonances. The coefficient for this term also has the opposite sign for most binaries where $R_P<9\bar{r}$, which translates the separatrix horizontally in phase space.   We have chosen  parameters such that $M_q=M_P/2$, $e=0.2$, $R_P=0.01\bar{r}$, and $(B_S-A_S)/C_S=0.015$. 

In the lower panel of Figure \ref{Figure:Surface_spinorbit}, we show a surface of section for the same orbital and separation parameters as for the schematic in the upper panel under the full evolution defined by the Hamiltonian given by Equation \ref{Equation:SpinOrbitHamiltonian}. Similarly to the prograde surface of section presented in  Figure \ref{Figure:ProgradeSurface}, we color code the points by the initial conditions of the trajectory. Trajectories that start below, in between and above the separatrix for the 1:1 spin orbit resonance are colored with turqouise, indigo and light blue points. We have chosen parameters such that the  1:1 and 3:2 resonances are overlapping. As can be seen in the surface of section, chaotic dynamics engulf the regions surrounding the nominal locations   of the two upper separatrices. Since the 1:2 resonance has a much smaller width than the other two, trajectories that start inside this resonance stay there. Smaller higher order resonances appear in  the regions between the $3:2$ and $1:1$ resonances located around $\alpha=\pi/2$ and $3\pi/2$.

 \citet{Nadoushan2015} demonstrated that increasing  the asphericity and eccentricity increased the width of the resonant regions numerically,  in good agreement with the analytic theory. They quantified  the effects of both parameters by computing surfaces of section, keeping the eccentricity constant and increasing the asphericity (Figure 7), and keeping the asphericity constant and increasing the eccentricity (Figure 8). Our numerical values of eccentrity and asphericity are most directly comparable to Panels B and C in Figure 8 of \citet{Nadoushan2015}, and the same higher order resonances appear in both figures. 
   
\section{Retrograde Spin-Spin Coupling}\label{section:retrograde}

In this section, we consider the spin dynamics of the satellite governed by the final term in the Hamiltonian given by Equation \ref{Equation:Hamiltonian_original}. We consider only the contribution to the torque from the last term in the equation  which has the form,

\begin{equation}
    \mathcal{H}_{ret}=\frac{\mathcal{A}^2}{2}-\frac{105GM_q\epsilon_P^2 }{16r^3}\bigg(\frac{B_S-A_S}{C_S}\bigg)\cos(4f-2(\alpha+\phi))\, , 
\end{equation}
By performing a trigonometric substitution, transforming from true to mean anomaly using Equations  \ref{Equation:cosf} and \ref{Equation:sinf},  and substituting the first order expansion in eccentricity for $1/r^5$ as defined in Equation  \ref{Equation:r5_expansion}, the expression  reduces, to first order in eccentricity, to,

\begin{equation}\label{Equation:Hamiltonian_retrogradespinspin}
    \mathcal{H}_{ret}=\frac{\mathcal{A}^2}{2}- \frac{105GM_q R_P^2}{16\bar{r}^5}\bigg(\frac{B_S-A_S}{C_S}\bigg)\left(\cos
   (4 M-(2 \alpha +2 \phi ))+e\frac{13}{2}  \cos ( 5 M-(2 \alpha+2 \phi) )-e\frac{3}{2}
    \cos (3M-(2 \alpha  +2 \phi) )\right)\, . 
\end{equation}
Interestingly, every term in this Hamiltonian depends on the spin of the primary, the spin of the satellite and the orbital motion. Therefore, the interpretation of each of these terms is slightly less straightforward as was the case for the prograde spin-spin coupling terms and the spin-orbit coupling terms. However, by performing similar contact transformations, it can be shown that the resulting relations between the spin of the satellite and the spin of the primary and mean motions are straightforward. 

\begin{figure}
\begin{center}
\includegraphics[scale=0.35 ,angle=0]{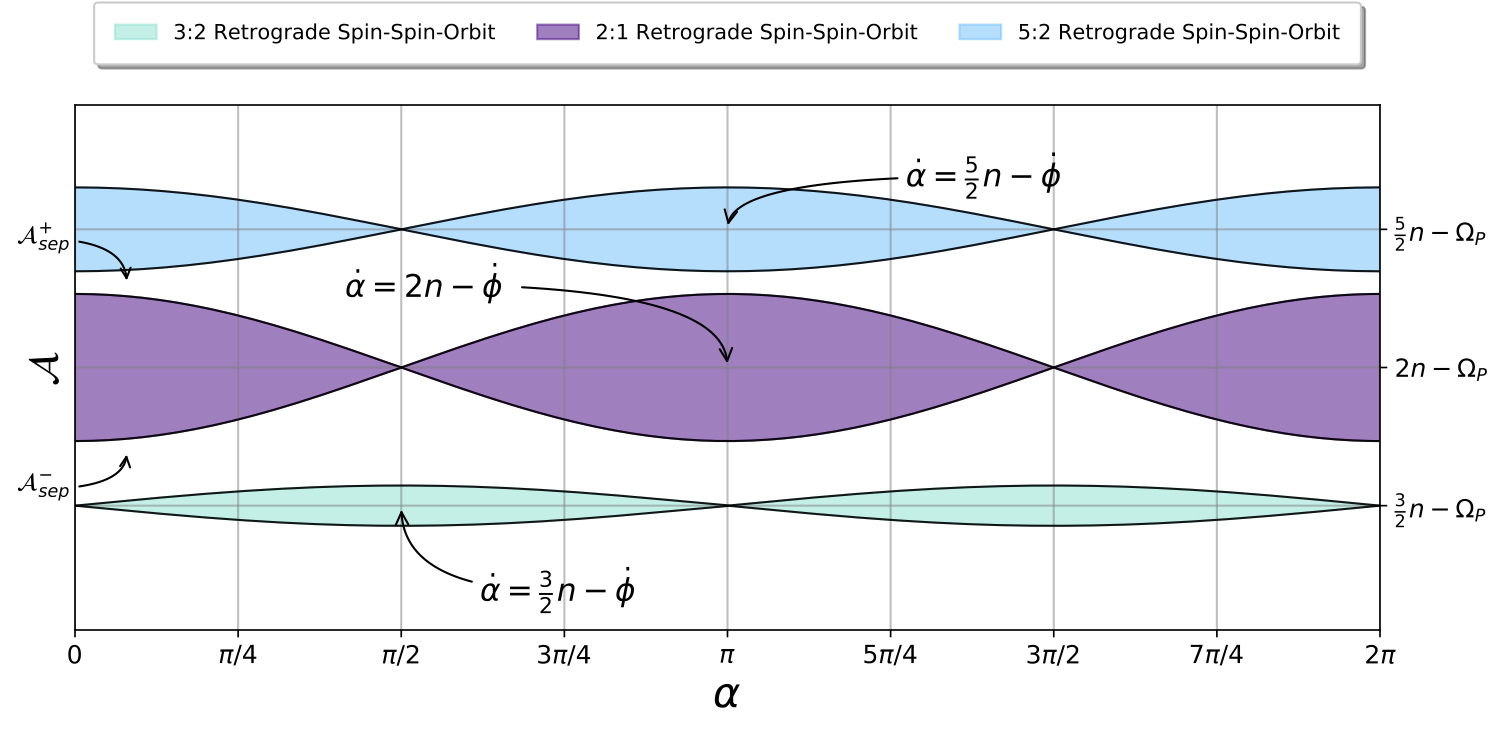}
\includegraphics[scale=0.35 ,angle=0]{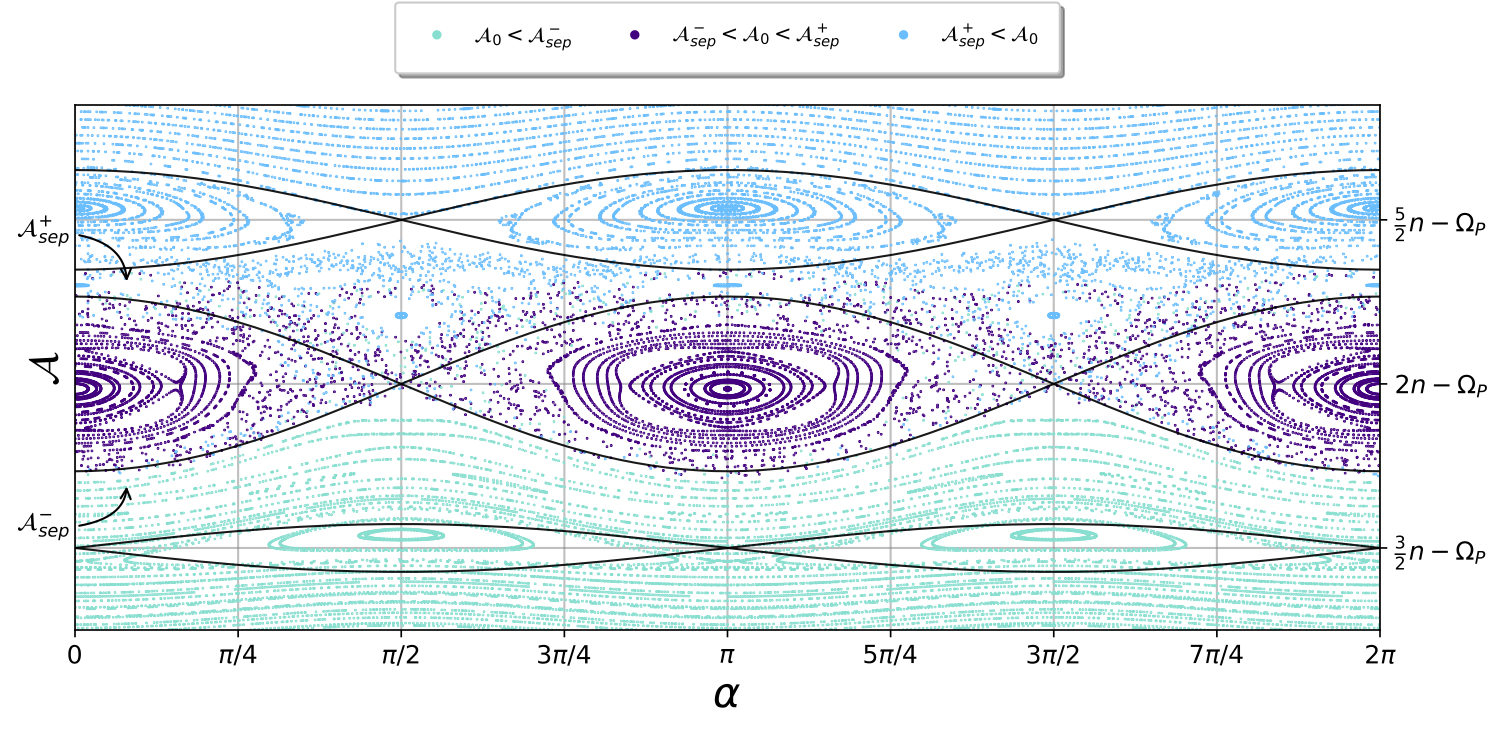}
\caption{The dynamics governed by  the retrograde spin-spin coupling term in the Hamiltonian. The top panel shows a schematic phase space diagram, where the black lines indicate the separatrices of the Hamiltonians arising from the three contact transformations corresponding to $\dot{\alpha}=2n-\dot{\phi}$, $\dot{\alpha}=\frac52n-\dot{\phi}$, and $\dot{\alpha}=\frac32n-\dot{\phi}$. The lower panel shows the surface of section for the Hamiltonian given by Equation \ref{Equation:Hamiltonian_retrogradespinspin}. The colors of the points indicate the intial conditions, similar to Figures \ref{Figure:ProgradeSurface} and \ref{Figure:Surface_spinorbit}.  The binary parameters are  $M_q=M_P/20$, $e=0.05$, $\epsilon_P=.3$, and the satellite has an aspect ratio $a/b\sim 2$. Although the resonances are not overlapping, chaotic mixing occurs close to the nominal locations of the separatrices of the upper two resonances.}\label{Figure:Retrogradespinspin}
\end{center}
\end{figure}

The zeroth order term in eccentricity corresponds to  a 2:1 resonance between the retrograde spin-spin coupling of the primary and satellite, and the mean motion. 
This Hamiltonian is centered at $\mathcal{A}=(2n-\Omega_P)$ and has a resonance half-width of $2\sqrt{ \frac{105GM_q R_P^2}{16\bar{r}^5}\big(\frac{B_S-A_S}{C_S}\big)}$.  Therefore, in the limit where $\Omega_P=0$, this is the 2:1 spin orbit resonance. Generally when the spins of the primary and satellite are well separated from the mean motion, this term is the 1:1 retrograde spin-spin coupling of the primary and satellite. The second 5:2 term is a resonance centered at $\mathcal{A}=(\frac{5n}{2}-\Omega_P)$ with a resonance half-width of $2\sqrt{\frac{1365GM_q R_P^2e}{32\bar{r}^5}\big(\frac{B_S-A_S}{C_S}\big)}$.  Therefore, this is either the  5:2 spin-orbit resonance or retrograde spin-spin resonance. The final term is the 3:2 term and is centered at $\mathcal{A}=(\frac{3n}{2}-\Omega_P)$ with a resonance half-width of $2\sqrt{\frac{315GM_q R_P^2e}{32\bar{r}^5}\big(\frac{B_S-A_S}{C_S}\big)}$, and represents  the 3:2 spin-orbit or retrograde spin-spin term. 

To evaluate the criteria for resonance overlap, we will only consider the of the 2:1 and 5:2 resonance, since the width of the 5:2 resonance is greater than that of the 3:2. The criteria for the onset of chaos will be,

\begin{equation}
    \frac{n}{2}<2\sqrt{ \frac{105GM_q R_P^2}{16\bar{r}^5}\bigg(\frac{B_S-A_S}{C_S}\bigg)}+2\sqrt{\frac{1365GM_q R_P^2e}{32\bar{r}^5}\bigg(\frac{B_S-A_S}{C_S}\bigg)}\, ,
\end{equation}
which can be simplified to,

\begin{equation}\label{eq:chaos_criteria_spinspinretrograde}
    \frac{\bar{r}}{R_P}<\sqrt{105} \sqrt{\frac{M_q}{M_P+M_S}}\sqrt{\bigg(\frac{B_S-A_S}{C_S}\bigg)}(1+\sqrt{\frac{13e}{2}})\,.
\end{equation}
As in the case of prograde spin-spin resonances, the fraction of phase space covered by the chaotic band would be zero as $e\rightarrow0$. The criteria for the upper resonance to completely overlap with the central resonance is,

\begin{equation}
    \frac{2}{13}\bigg(\frac{n}{4} \bigg(\frac{105GM_q R_P^2}{16\bar{r}^5}\frac{B_S-A_S}{C_S}\bigg)^{-1/2}+1\bigg)^2<e\,.
\end{equation}

The upper panel of Figure \ref{Figure:Retrogradespinspin} shows a schematic phase space diagram, with the separatrices of the three resonances shown in black. The light blue and indigo regions show the librational regions of the 5:2 resonance and the 2:1 resonance. The negative coefficient of the 3:2 resonance translates the separatrix horizontally in phase space by $\pi/2$, and the resonance has a smaller width than the other two. We have chosen to use a more moderately shaped primary where $M_q=M_P/20$, $e=0.05$, $\epsilon_P=.3$, and the satellite has an aspect ratio $a/b\sim 2$. We chose a more moderate system than for the prograde system to demonstrate that, since the widths of the resonances are larger, there will be resonance overlap for a larger variety of  systems.  The lower panel of Figure \ref{Figure:Retrogradespinspin} shows a surface of section for the  Hamiltonian given by Equation \ref{Equation:Hamiltonian_retrogradespinspin}. The separatrices of the upper two resonances are close to, but not  overlapping. As can be seen, the higher order resonances have caused chaotic dynamics in the regions closest to the separatrices, and there is phase space mixing where light blue points that started in the 5:2 state ``hop" into the 2:1 resonance. Although even the 3:2 term is more separated, the shape of the librations have been distorted. 

\begin{figure}
\begin{center}
\includegraphics[scale=0.35 ,angle=0]{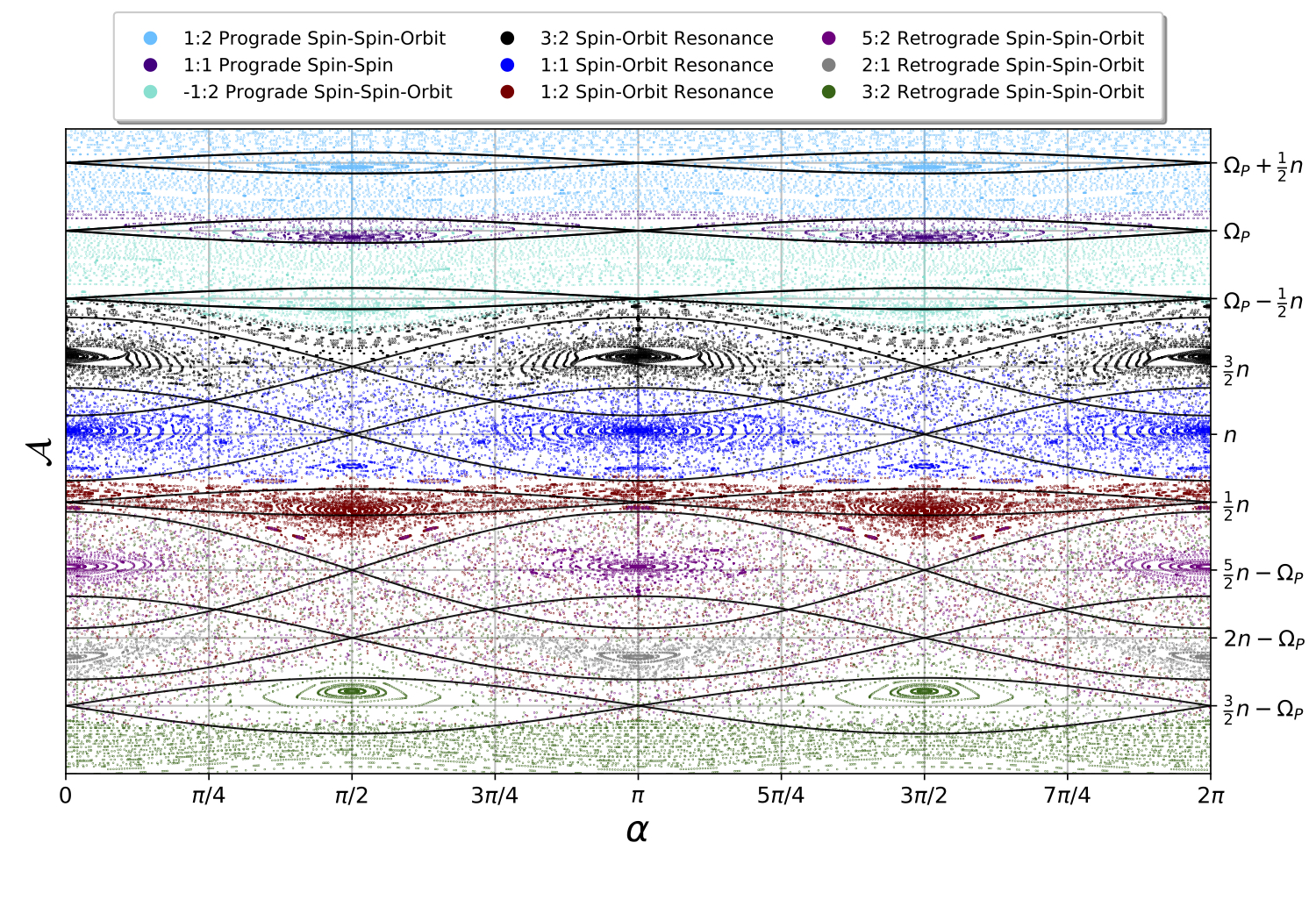}
\caption{A surface of section for the full Hamiltonian given by Equation \ref{Equation:FullHammy}. The colors of the points correspond to which resonances their trajectory is initialized in. The primary is spinning in the  same direction as the mean motion, so that the resonances are somewhat separated. Many of the resonances still overlap, and there is clear chaos in the system, specifically around the spin-orbit resonances and the retrograde spin-spin resonances. The parameters of the binary are $\Omega_P=5n/2$,    $M_Q=M_P$,  $e=0.3$,  $R_P=0.6\bar{r}$, and  $(B_S-A_S)/C_S=0.01$.}\label{Figure:fullevolution}
\end{center}
\end{figure}
\section{Overlap of All Resonances and the Onset of Chaos}\label{section:overlap}

In this section, we consider the evolution of the full Hamiltonian presented by Equation \ref{Equation:Hamiltonian_original}. Combining all of the substitutions performed in the previous three sections, this Hamiltonian can be re-written as,

\begin{equation}\label{Equation:FullHammy}
\begin{split}
        \mathcal{H}=\frac{\mathcal{A}^2}{2}-\frac{3G }{8 \bar{r}^3}\bigg(\frac{B_S-A_S}{C_S}\bigg)\bigg(\big(2M_P+ (4+5\epsilon_p^2)M_q\big)\cos (2 M-2 \alpha )\\- \frac{e}{2}\big(2M_P+ (4-5\epsilon_p^2)M_q\big)\cos (M-2
   \alpha )+\frac{e}{2}\big(14M_P+ (28+45\epsilon_p^2)M_q\big)\cos (3 M-2 \alpha )\bigg)  \\ -\frac{9GM_q R_P^2 }{16\bar{r}^5}\bigg(\frac{B_S-A_S}{C_S}\bigg)\bigg(\cos(2\alpha-2\phi)+\frac52e\cos(2\alpha-2\phi-M)+\frac52e\cos(2\alpha-2\phi+M)\bigg)\\- \frac{105GM_q R_P^2}{16\bar{r}^5}\bigg(\frac{B_S-A_S}{C_S}\bigg)\left(\cos
   (4 M-(2 \alpha +2 \phi ))+e\frac{13}{2}  \cos ( 5 M-(2 \alpha+2 \phi) )-e\frac{3}{2}
    \cos (3M-(2 \alpha  +2 \phi) )\right)
\end{split}
\end{equation}
 This Hamiltonian may be directly compared to the Hamiltonian presented in Equation 9 of \citet{Nadoushan2016}. This Hamiltonian has the same harmonics, with the exception of the terms resulting from the spin of the primary only, and those to higher order in the asphericity of the Satellite. 

 In Figure \ref{Figure:fullevolution}, we show a surface of section for the evolution of the system under the action of the full Hamiltonian.  We take the surface of section every super period, or at a period that corresponds to the least common multiple of all of the periods. The primary is spinning in the same sense as the mean motion, and $\Omega_P=5n/2$\footnote{We have chosen $\Omega_P$ to be a rational multiple of the mean motion purely in order to capture the dynamics in the surface of section.}.   We have chosen a system where $M_Q=M_P/2$, in order to maximize the strength of the spin-spin resonances with respect to the spin-orbit resonances, which will dominate in less dramatic systems. The eccentricity is $e=0.3$,  the primary radius is $0.6\bar{r}$, and the satellite has $(B_S-A_S)/C_S=0.01$. The points are color coded based on the resonance that they begin in, and many of the resonances overlap. Although some resonant structures still appear, the system is almost entirely chaotic, and there is clear phase space mixing and resonance jumping. There are some curious features that appear in the surface of section. Specifically, the libration in the $3:2$ spin-orbit resonance appear to have spiral structure. However, the structure is not actually spiral, and it only looks this way due to the existence of higher-order resonances embedded within the separatrix itself.

The Hamiltonian presented in Equation \ref{Equation:FullHammy} can be loosely compared with the  Chirikov standard map to yield an approximate condition for chaos in a  system. This connection technically applies only in the special case where all the resonant harmonics are spaced equidistantly in frequency and have approximately similar widths. While these assumptions can be relaxed to create a general rule for the onset of chaos, this connection is illuminating from a heuristic point of view. We suppose that phase space looks like an equidistant stack of resonances and that the strength, $c$, of the individual resonances are all comparable. Each harmonic then, upon averaging, is describe by a pendulum like Hamiltonian,

\begin{equation}
    \mathcal{H}=\frac12\mathcal{A}^2+c\cos(\alpha)\,.
\end{equation}
The total Hamiltonian is then akin to

\begin{equation}
    \mathcal{H}=\frac12\mathcal{A}^2+c\bigg(\cos(\alpha)+\cos(\alpha-n/2t)+\cos(\alpha-nt)+...+\cos(\alpha+n/2t)+\cos(\alpha+nt)+...\bigg)\,,
\end{equation}
where each term corresponds to a harmonic in Equation \ref{Equation:FullHammy} that is a linear combination of $\alpha$, $n$ and $\Omega_P$.
Because the spin-orbit and spin-spin problem has a total of 9 harmonics, we can write

\begin{equation}
    \mathcal{H}=\frac12\mathcal{A}^2+c'\bigg(\frac{1}{2\pi}+\frac{1}{\pi}\sum_{i=1}^9\cos( in/2t)\bigg)\cos(\alpha)\,,
\end{equation}
where $c'=2\pi c$. Recall that the Fourier series of a delta function is given by,

\begin{equation}
    \delta = \frac{1}{2\pi}+\frac{1}{\pi}\sum_{i=1}^\infty\cos(in/2t)\,.
\end{equation}
If we define the angle $\theta=n/2t$, then this definition of the delta function is $2\pi$ periodic in $\theta$. If we make the approximation that $9\sim\infty$\footnote{In fact, this approximation is better than it seems, as long as we only limit the action-domain to the overlapped region.}, then

\begin{equation}
    \mathcal{H}=\frac12\mathcal{A}^2+c'\cos(\alpha)\delta\,.
\end{equation}

This Hamiltonian gives a mapping with,

\begin{equation}
    \Delta \alpha=\frac{\partial\mathcal{H}}{\partial \mathcal{A}}=\mathcal{A}\,,
\end{equation}
and
\begin{equation}
    \Delta \mathcal{A}=-\frac{\partial\mathcal{H}}{\partial \alpha}=c\sin(\alpha)\,.
\end{equation}

This gives the mapping

\begin{equation}
    \begin{cases}
    \mathcal{A}_{i+1}=\mathcal{A}_{i}+c\sin(\alpha_i),& \\
    \alpha_{i+1}=\alpha_i+\mathcal{A}_{i+1},              & 
\end{cases}\,,
\end{equation}
This is the Chirikov standard map, which shows chaos for $c>1$ and has a well-characterized diffusion coefficient $\mathcal{D}_D$,
\begin{equation}
    \mathcal{D}_D\sim c^2/2, \, \text{for} \,\, c>>1\,.
\end{equation}

We can roughly apply this criteria to our Hamiltonian system in Equation \ref{Equation:FullHammy}.  If we consider  only the widths of the resonances to leading order in eccentricity, then the system will be chaotic if

\begin{equation}
    \max \bigg[ \frac{3G }{8 \bar{r}^3}\big(\frac{B_S-A_S}{C_S}\big)\bigg(\big(2M_P+ (4+5\epsilon_p^2)M_q\big) \, , \, \frac{3M_q \epsilon_P^2 }{2} \, ,\, \frac{35M_q \epsilon_P^2}{2}\bigg)\bigg]>1\,.
\end{equation}
Although this is a useful approximate criteria, and an elegant connection to the Chirikov standard map, we stress that the criteria for chaos for each of the individual resonances derived in \S\ref{section:prograde}-\ref{section:retrograde} are more rigorous.


\section{ 3-Dimensional Motion: Obliquity Variations}

In this section, we investigate the obliquity evolution of a permanently deformed satellite in a spin-spin resonance. We extend the geometry of the planar case to include the inclination $i$ and node $\Omega$ of the satellite, and evaluate the obliquity variations that the satellite will undergo in the individual resonances. The node $\Omega$ is measured in the x-y plane, and indicates where the satellite tilts out of the plane with inclination, $i$.  The original, planar orientation of the satellite, which was denoted by $\alpha$, is now generalized to the orientation about the new spin axis.  The 3-dimensional positions of the first point mass in the satellite are given by

\begin{equation}
    x_1=r\cos (f)+\epsilon_sr\big(\cos(\alpha)\cos(\Omega)-\cos(i)\sin(\alpha)\sin(\Omega)\big)\,,
\end{equation}
\begin{equation}
     y_1 = r\sin (f)+\epsilon_sr\big(\cos(i)\sin(\alpha)\cos(\Omega)+\sin(\Omega)\cos(\alpha)\big)\,,
\end{equation}
and
\begin{equation}
     z_1 = \epsilon_sr\sin(i) \sin(\alpha)\,,
\end{equation}
\begin{figure}
\begin{center}
\includegraphics[scale=0.7,angle=0]{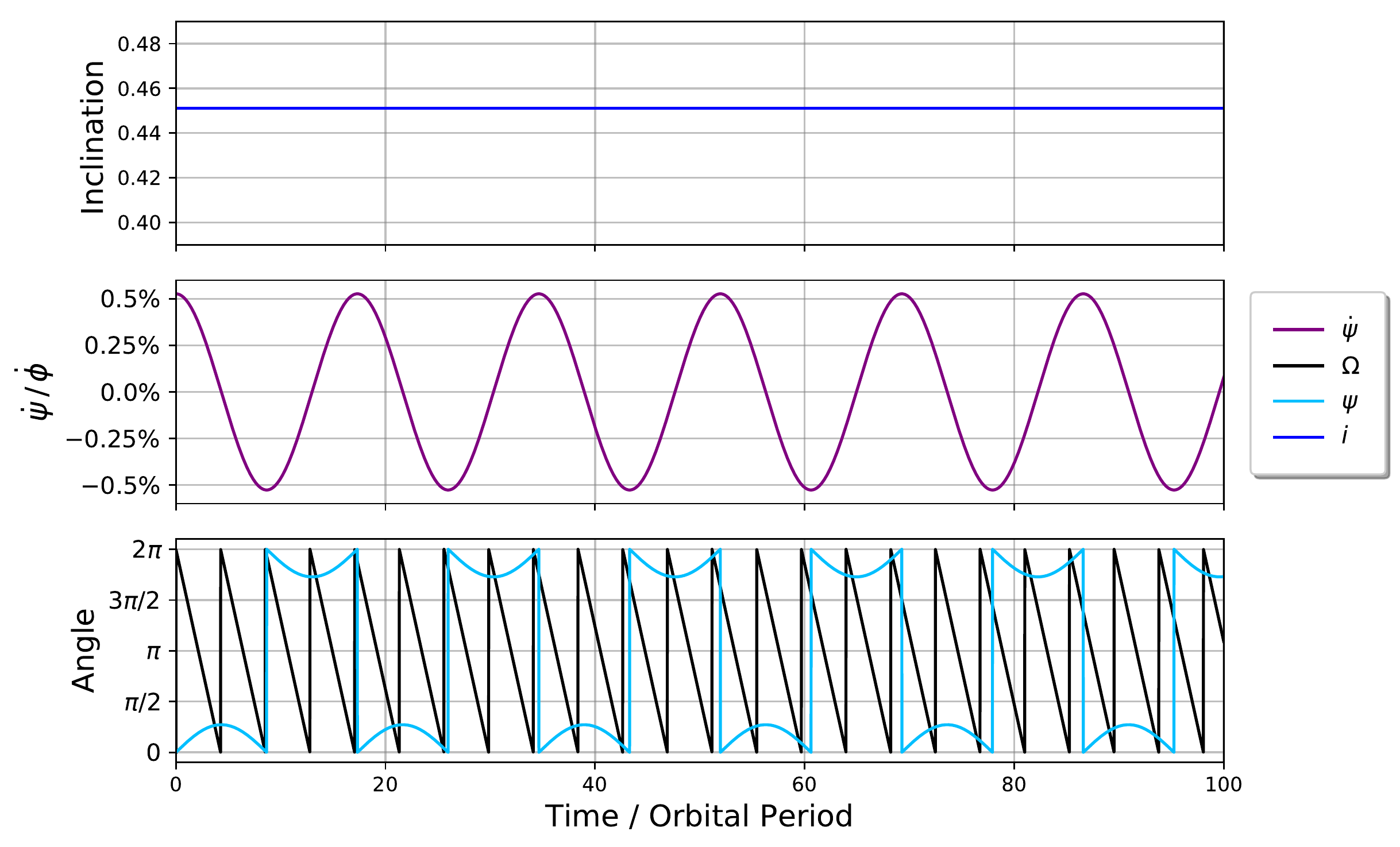} 
\caption{ The obliquity evolution of a satellite in the prograde spin-spin resonance, whose spin axis is initially inclined with respect to the orbital plane. The evolution of the inclination,  action $\Psi$ and node $\Omega$ and angle $\psi=\alpha-\phi+\Omega$ are shown in the top, middle and bottom panels respectively. $\Psi$ undergoes librations with an amplitude  $\sim 0.5\%$ of the spin frequency of the primary, while the angle $\psi$ librates close to the equilibrium point defined by $[\psi]=0$. The spin vector of the satellite   precesses, and the node $\Omega$  circulates through $0$ to $2\pi$ while the inclination remains constant. }\label{Fig:ObliquityEvolution}
\end{center}
\end{figure}
 while the second point mass is defined analogously to Equation \ref{eq:x2y2}, and $z_2=-z_1$. The two degree of freedom Hamiltonian may be evaluated by summing the potential energy of each point mass pair and the kinetic energy of the satellite, and expanding to second order in $\cos(i)$, $\epsilon_P$ and $\epsilon_S$. Because we are only interested in the obliquity evolution and not the relative strengths of the resonances, we do not include the central mass of the primary in the sum, and the Hamiltonian  has the form,

\begin{equation}\label{Eq:obliquity_hammy_full}
\begin{split}
        \mathcal{H}_{obl}=\frac{\mathcal{A}^2}{2}-\frac{G M_q}{64r}\frac{1}{R_S^2}\bigg(\frac{B_S-A_S}{C_S}\bigg)\bigg[ 128+16\epsilon_S^2\big(3 \cos^2(i)-1\big)+32 \epsilon_P^2+36\epsilon_P^2\epsilon_S^2\big(3 \cos^2(i)-1\big)+\\
        \bigg(12 \epsilon_S^2(4 + 9 \epsilon_P^2)  \sin^2(i)  \bigg) \cos(2\alpha)+\bigg(
        96\epsilon_P^2 + 60\epsilon_P^2\epsilon_S^2 \big(1 + \cos^2(i) - 2 \sin^2(i)\big)\bigg) \cos(
  2 f - 2 \phi)+\\
  \bigg(90\epsilon_P^2 \epsilon_S^2\sin^2(i)\bigg) \bigg(\cos(2 f - 2 \alpha - 2 \phi)+\cos(2 f +2 \alpha - 2 \phi)\bigg)+\\\bigg(12 \epsilon_S^2(4 + 5 \epsilon_P^2)  \sin^2(i) \bigg) \cos(
  2 f - 2\Omega)+\bigg( 6 \epsilon_S^2 (4 + 5 \epsilon_P^2)  \big(1 + \cos(i)\big)^2\bigg) \cos(
  2 f - 2\alpha - 2\Omega)+\\\bigg(
 105\epsilon_P^2 \epsilon_S^2 (1 - \cos(i))^2\bigg)  \cos(
  4 f + 2\alpha - 2\phi - 2\Omega)+\bigg(18 \epsilon_P^2 \epsilon_S^2 \sin^2(i)\bigg) \cos(2 \phi - 2 \Omega)
  +\\
  \bigg(6 \epsilon_S^2(4 + 5 \epsilon_P^2)   (1 - \cos(i))^2 \bigg)\cos(
     2 f + 2\alpha - 2\Omega) + \\
  \bigg( 9 \epsilon_S^2 \epsilon_P^2 (1 - \cos(i))^2 \bigg)\cos(
        2 \alpha + 2\phi - 2\Omega) +\bigg( 9 \epsilon_S^2 \epsilon_P^2(1 + \cos(i))^2\bigg) \cos
         (2\alpha -2 \phi +2 \Omega)+\\
         \bigg(210 \epsilon_P^2 \epsilon_S^2  \sin^2(i)\bigg) \cos(
  4 f - 2 \phi - 2\Omega)+\bigg(105\epsilon_P^2 \epsilon_S^2 (1 + \cos(i))^2 \bigg) \cos(
  4 f - 2 \alpha - 2\phi - 2\Omega)
    \bigg]\, ,
\end{split}
\end{equation}

In the limiting case of zero obliquity where $i=0=\Omega$, Equation \ref{Eq:obliquity_hammy_full} reduces  to the planar Hamiltonian presented in Equation \ref{Equation:Hamiltonian_original}, without the central primary mass. 

While this Hamiltonian may be used to evaluate the entire obliquity evolution of the satellite, we specifically focus on the cases where the satellite is entrained within a prograde or retrograde spin-spin resonance, and resides at the resonant equilibrium point. 
For these cases, the rate of change of $\alpha$ is close to the spin of the primary and $|d \alpha/dt|\sim d\phi/dt$.  Using scaled versions of the Delaunay variables, $H=\cos(i)$ and $h=\Omega$, and resulting equations of motion, the rate of change of the node $\Omega$, is of the order,  $d \Omega/dt\sim GM_q\epsilon_P^2/(64r^3) (B_S-A_S)/C_S$, and therefore   $|d \alpha/dt|>>|d \Omega/dt|$. This condition essentially ensures that the $(\alpha, \mathcal{A})$ degree of freedom will behave in an adiabatic manner with respect to the obliquity evolution, characterized by the $(\Omega,\cos(i))$ degree of freedom. In fact, this assumption is robust, and only breaks down for very slow spin-rates of the primary, i.e., considerably slower than the orbital frequency of the satellite.   Therefore, as long as the primary is spinning faster than the precession rate, we may  utilize the adiabatic evolution of the precession to  examine the dynamical character of the $(\alpha, \mathcal{A})$ equilibrium point on the $(\Omega,\cos(i))$ phase-plane.

In order to evaluate the obliquity evolution in the prograde spin-spin resonance, we  employ a similar type-2 generating function to Section 3.1, to transform to an angle representative of the spin-spin resonance.  Defining the angle $\psi =\alpha-\phi+\Omega $ and action $\Psi = \mathcal{A}$, we transform the above Hamiltonian, which we  then integrate  over all primary orientations $\phi$ and points along the orbit $f$ through 0 to $2\pi$, normalized by $4\pi^2$. This yields a time averaged, transformed Hamiltonian for the prograde spin-spin resonance $\tilde{\mathcal{H}}_{pro}$, which is of the form,

\begin{equation}
\begin{split}
    \tilde{\mathcal{H}}_{pro} =\frac12 \bigg(\Psi-\frac{d\phi}{dt}+\frac{d\Omega}{dt}\bigg)^2-\frac{9G M_q\epsilon_P^2}{64r^3}\bigg(\frac{B_S-A_S}{C_S}\bigg)\big(  1 + \cos(i)\big)^2 \cos
         (2\psi)\\
         -\frac{G M_q}{16r}\frac{1}{R_S^2}\bigg(\frac{B_S-A_S}{C_S}\bigg)\bigg( 32+4\epsilon_S^2\big(3 \cos^2(i)-1\big)+8 \epsilon_P^2+9\epsilon_P^2\epsilon_S^2\big(3 \cos^2(i)-1\big)\bigg)
        \end{split}
\end{equation}

Because of the seperation of the timescales, for a given $(i,\Omega)$, we denote the equilibrium points of $\psi, \Psi$ as $[\psi]$ and $[\Psi]$. We  then Taylor expand the Hamiltonian around these equilibrium values, where $\bar{\Psi}=\Psi-[\Psi]$ and $\bar{\psi}=\psi-[\psi]$. The equilibrium points will be located where the spin of the satellite is roughly equal to the spin of the primary, with the phase modulation of the node (which  varies much slower than $\alpha$ and $\phi$), $[\Psi]=\dot{\phi}-\dot{\Omega}$, and at $[\psi]=\alpha-\phi+\Omega$. The  Hamiltonian expanded  around the equilibrium point to second order is given by, 

\begin{equation}\label{eq:taylor_expanded_obliquity}
    \tilde{\mathcal{H}}_{\rm{exp}}=\frac12 \bar{\Psi}^2+\frac{9G M_q\epsilon_P^2}{32r^3}\bigg(\frac{B_S-A_S}{C_S}\bigg)\big(  1 + \cos(i)\big)^2\bar{\psi}^2 \, .
\end{equation}

Because the coefficient of the $\bar{\psi}^2$ term  is positive definite,  the equilibrium point is elliptic (the Hamiltonian is analogous to a harmonic oscillator), and the resonance center is stable  $\forall(i,\Omega)$. This   proves that the adiabatic approximation holds for all obliquity\footnote{ Formally,  the resonance width will go to zero for inclinations $i=\pm i/2$}. Setting the system to reside at the resonance center, we can  therefore  treat  Equation  \ref{eq:taylor_expanded_obliquity}  as  an effectively one  degree-of-freedom  Hamiltonian,  for  inclination  and  node. At the resonance center, $\psi=0$, and the equations of motion for the Delaunay variables show that $d\cos(i)/dt\sim \rm{constant}$, while the node, $\Omega$, circulates. This demonstrates that the satellite will undergo precession while in the prograde spin-spin resonance. In Figure \ref{Fig:ObliquityEvolution}, we show the obliquity evolution for a satellite that begins inclined to the orbit plane and close to the prograde spin-spin resonance  over 100 orbital periods. While $\psi$ librates close to the equilibrium value of $[\psi]=0$, the spin vector of the satellite precesses. For this particular experiment, we computed the evolution for a primary with $\epsilon_P=0.1$, and  a satellite   with an aspect ratio of $\sim3.5:1$  on a circular orbit. The primary was initialized with a spin frequency that was 3 times that of the mean motion of the satellite, and the satellite had an initial inclination of $i_0\sim 25^\circ$ out of the orbital plane.   

 A directly analogous analysis for the term $\sim \cos(4f-2\alpha-2\phi-2\Omega)$ reveals the same conclusion for  the retrograde spin-spin resonance. This analysis reveals that the resonances themselves are vertically stable, and tumbling would likely require the primary to be significantly inclined with respect to the orbital plane, or mixing among the degrees of freedom i.e., chaotic motion.

\section{Applications and Discussion}\label{section:kleopatra}
\begin{figure}
\begin{center}
\includegraphics[scale=0.7,angle=0]{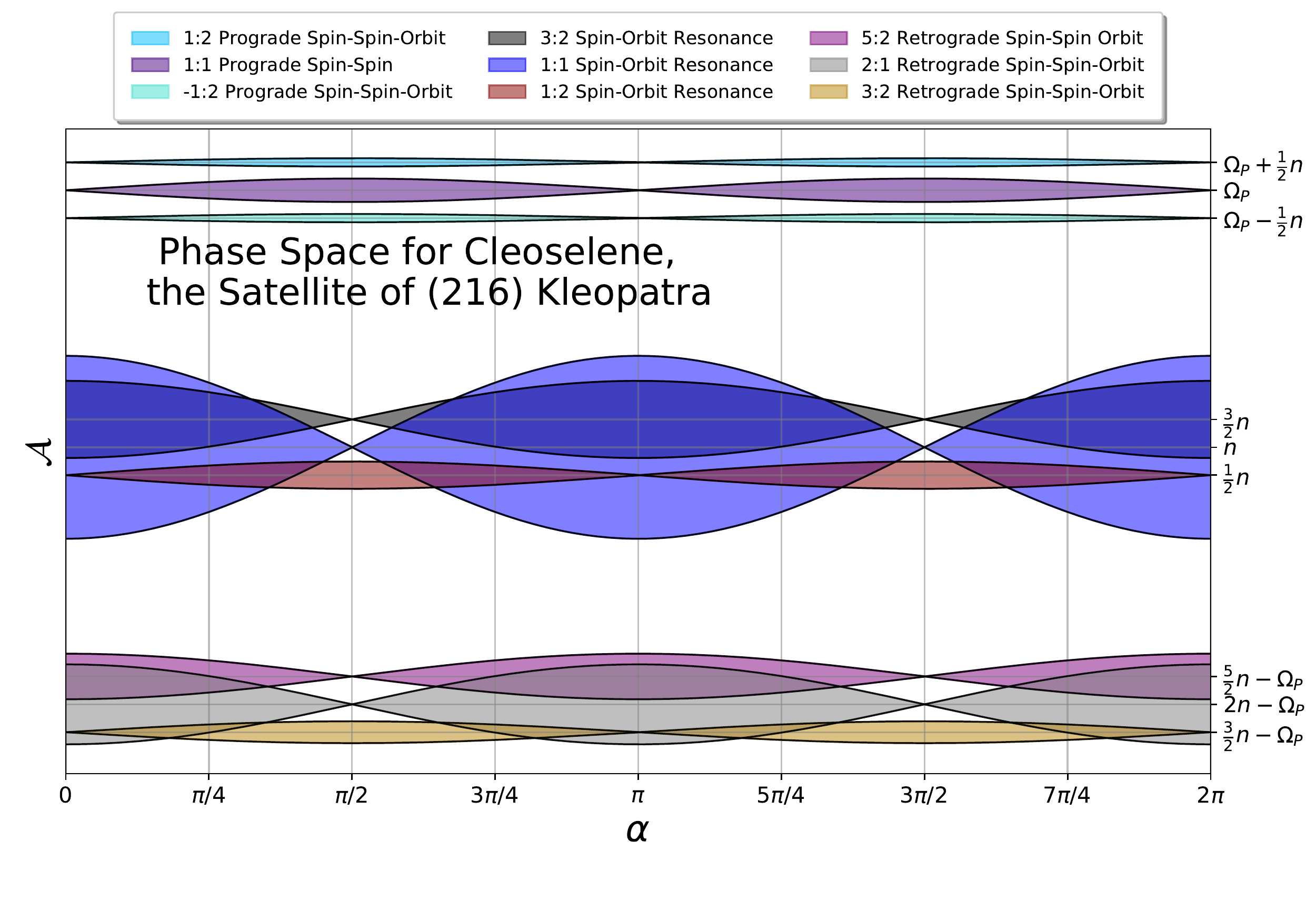} 
\caption{The phase space showing the nine resonances for the inner moon of  (216) Kleopatra, Cleoselene. We have used the parameters for the moon and for Kleopatra presented in \citet{Descamps2011} and \citet{Shepard2018}, where $R_P=138$ km and $M_P=4.65\times10^{18}$kg. The period of Kleopatra is $5.39$ hours and the period of the moon is $1.24\pm0.02$ days (at a semi major axis $\bar{r}=454$ km). With these parameters $n=2\pi\, 0.80 \,\rm{days} ^{-1}$ and $\Omega_P=2\pi\, 4.45 \, \rm{days}^{-1}$. We assume that the moon has an asphericity like Nix, with $(B_S-A_S)/C_S=0.63$, modified by the cosine of the inclination. The spin-orbit and retrograde spin-spin resonances are overlapping. If Cleoselene's spin rate is in the vicinity of these resonances, it should be undergoing chaotic rotation. }\label{Fig:PhasespaceKleo}
\end{center}
\end{figure}

\begin{figure}
\begin{center}
\includegraphics[scale=0.7,angle=0]{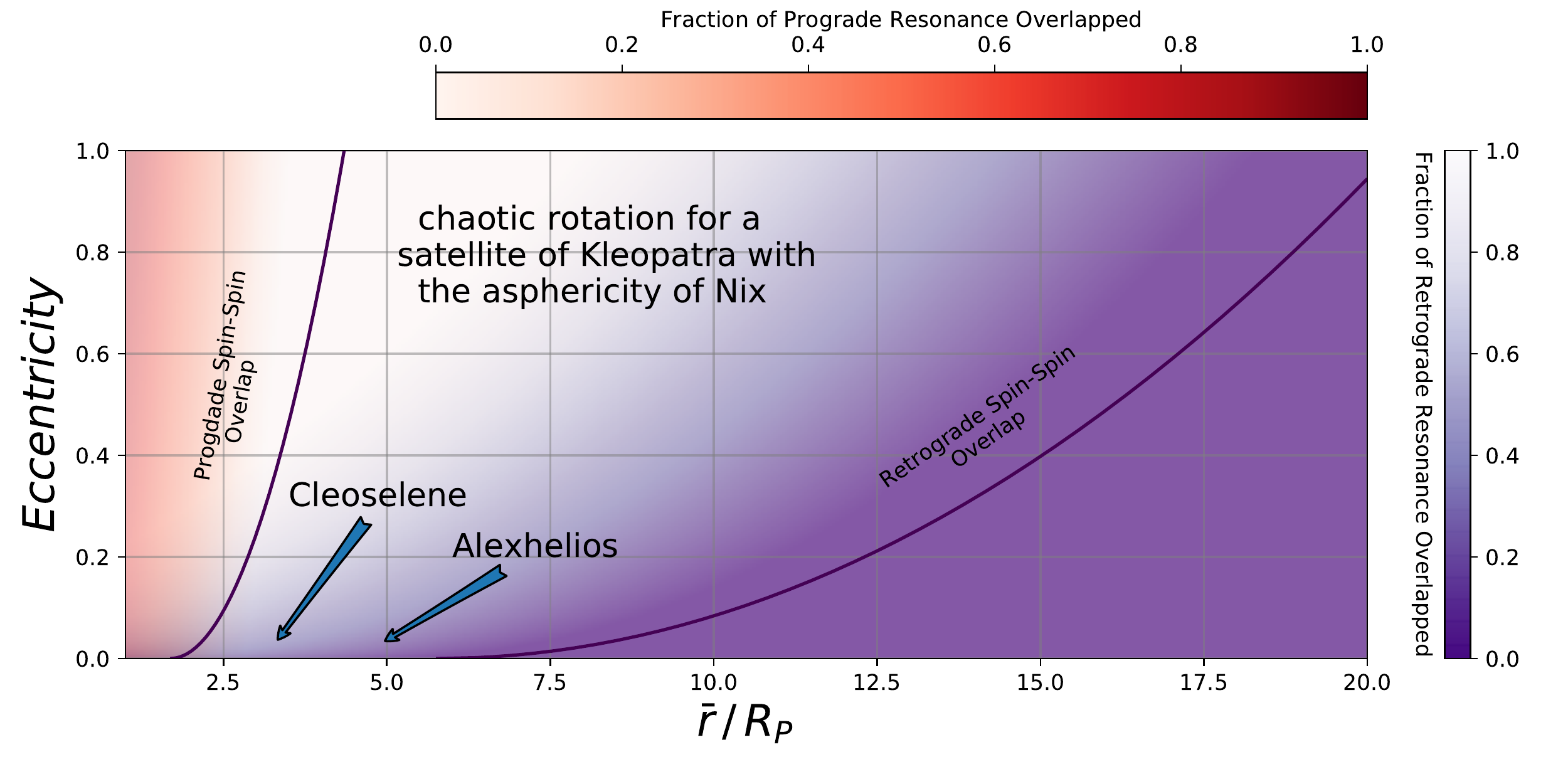} 
 \caption{The eccentricity and orbital seperation for a \textit{hypothetical} satellite of (216) Kleopatra that would result in chaotic rotation from prograde or retrograde spin-spin overlap. The position of the two known satellites, Cleoselene and Alexhelios, are marked on the figure. The two contours  indicate where Equations \ref{eq:chaos_criteria_spinspinprograde} and \ref{eq:chaos_criteria_spinspinretrograde} are satisfied, where the prograde and retrograde spin-spin resonances will overlap. The red and purple color gradients indicate the fractional area of the central resonance that is overlapping with a secondary resonance. For this entire parameter space, the spin-orbit resonances are overlapping. While these criteria are calculated for (216) Kleopatra, they are also generally applicable to similarly deformed contact binaries in the Solar System.}\label{Fig:Kleopatra}
\end{center}
\end{figure}
In this section, we discuss some of the possible applications of this framework to real astrophysical systems. In particular, we focus on the triple system of the asteroid (216) Kleopatra, as a case study, in order to demonstrate that satellites of other contact binaries may be in chaotic rotational states driven in part by spin-spin coupling. The two moons of Kleopatra, Cleoselene and Alexhelios, were discovered by \citet{Descamps2011}, and have orbital periods of $1.24\pm0.02$ and $2.32\pm0.02$ days at semi-major axis of 454 and 678 km, respectively. These measured periods imply a system mass of $\sim 4.65\times10^{18}$ kg (for concise descriptions of these parameters see Table 2 in \citet{Descamps2011} and Tables 1 and 2 in \citet{Hirabayashi2014}). \citet{Shepard2018} presented the most recent shape model of the primary of Kleopatra, with    a spin period of $5.39$ hours, and a shape of $276\times94\times 78$ km. The diameters of Cleoselene and Alexhelios are $6.9\pm1.6$ km and $8.9\pm1.6$ km respectively, although their shape and aspect ratio are unconstrained. 

Several authors have investigated the stability of orbits around Kleopatra, motivated by the discovery of the satellites. \citet{Descamps2011} hypothesized that ``as to the origin of Kleopatra’s companions, they could be a by-product of the spinning-up process leading to mass shedding in orbit." \citet{Yu2012} found that equilibrium orbits in the vicinity of Kleopatra were non-linearly unstable, which implied that the quasi-stationary orbits of satellites of Kleopatra should migrate during their evolution. \citet{Yu2013} used Monte Carlo dynamical simulations to identify a large set of resonant orbits around Kleopatra, and found that the unstable boundary was consistent with the location of the two satellites. However, \citet{Hirabayashi2014} found that at the $\sim5.4$ hour spin period of Kleopatra, surface shedding was unlikely to have occured, and postulated that the satellites may be the result of another process such as the re-accumulation of an impact-generated debris disk.  \citet{Chanut2015MNRAS} found that the minimum orbital distance for prograde and retrograde circular orbits  around Kleopatra that would  not result in an impact were 300 km ($\sim2.17R_P$) and 240 km ($\sim1.73R_P$), respectively. However, they found that increasing the orbital eccentricity of test particles increased their stability in ranges of eccentricity from $e=0.0-0.3$.

In order to model this system, due to Kleopatra's ``dog-bone" shape, we assume that $M_q=M_P/2$. This is a reasonable estimate for the quadrupole mass, which is itself simply a heuristic. Relaxing this assumption will only decrease the widths of the spin-spin resonances by a factor of $\sqrt{M_q}$.  We  assume that the primary radius is $R_P=138$ km and corresponds to the long axis in the best shape model. Since the smallest objects characterized in the Solar System have very aspeherical shapes,  we assume that the satellite has an asphericity $(B_S-A_S)/C_S\sim0.2$. This is a modest assumption, as the asphericity  of the satellite of Pluto, Nix was measured to be $(B_S-A_S)/C_=0.63$   \citep{Showalter2015,Correia2015}. For the purposes of this, we assume that the satellite has a very small eccentricity $e=0.05$, since \citet{Chanut2015MNRAS} found that close orbits to Kleopatra were more stable if they had some  eccentricity. The resonances will still overlap if their eccentricity is as small as $e=0.01$.  The satellite has an  inclination of $49^\circ$, although \citet{Hirabayashi2014}  assumed that the orbital planes of the satellites were parallel to the equatorial plane of (216) Kleopatra for the purpose of dynamical modeling. In light of the obliquity analysis in the previous section, the inclination of the satellite may introduce precession if it is in the spin-spin resonances. In the scope of this paper we are restricting our analysis to 2-dimensions. Future observations that could characterize the shape of the satellite would motivate more detailed 3-dimensional dynamical simulations. 

In Figure \ref{Fig:PhasespaceKleo} we show the phase space and resonances  for the parameters listed above for the moon of Kleopatra, Cleoselene. The three spin-orbit resonances and the three retrograde spin-spin resonances are overlapping. Therefore, based on  the spin period of the satellite, Cleoselene could feasibly be in any of these resonances, and exhibiting chaotic rotation. Future observations that could  reveal the shape and spin period of the satellite would also inform whether or not it is exhibiting chaotic rotation.

In Figure \ref{Fig:Kleopatra}, we show the parameter space of orbital seperation and eccentricity for a \textit{hypothetical} satellite of Kleopatra that would result in chaotic rotation from prograde or retrograde spin-spin overlap. The two contours on the figure indicate where the retrograde and the prograde spin-spin resonances will begin to overlap, corresponding to satisfying Equations \ref{eq:chaos_criteria_spinspinprograde} for the prograde case and Equation \ref{eq:chaos_criteria_spinspinretrograde} for the retrograde case. The two color-bars indicate the fractional area of the central resonance that is overlapped. When the eccentricity  approaches zero, the fractional area of the overlapping resonance goes to zero. This is because  for very small, nonzero eccentricities, the resonances  technically still exist, but their widths approach zero. Due to the findings in \citet{Chanut2015MNRAS}, the regions in this plot where the eccentricity approaches zero and the separation is less than $\sim 2R_P$ will result in unstable orbits. In order to calculate the fractional area of the central resonance that is overlapping, for each set of separation and eccentricity in Figure \ref{Fig:Kleopatra}, we calculate separatrices for the corresponding resonances and numerically integrate the area between the separatrices for the central resonance and overlapping region between the two resonances in a region corresponding to a domain with length of $\pi/2$ around the fixed point. 

For all of the regions of parameter space shown in the plot, the spin-orbit resonances overlap and Equation \ref{eq:chaos_criteria_spinorbit} is satisfied. In general, for a body as aspherical as Nix, the spin-orbit resonances will overlap out to orbital separations $\bar{r}>1000R_P$. Generally, bodies that have spin-spin resonances overlapping will also have spin-orbit resonances overlapping. However, for satellites of contact binaries, the orbital period and spin period of the primary will be well separated, as is the case in Figure \ref{Fig:PhasespaceKleo}, and the spin-spin resonances will not overlap with the spin-orbit resonances. The position of the two satellites, Cleoselene and Alexhelios, in this parameter space are indicated. Although the asphericity of these satellites are unconstrained, if they are similar to Nix in shape, then they could both lie within the chaotic retrograde spin-spin state.

These spin-spin resonances are a weak effect for most bodies in the Solar System. However, for small satellites around contact binaries where the primary exhibits a strong quadrupole moment and $M_q\sim M_P/2$, these spin-spin resonances may be an important driver of chaotic dynamics. Although Figure \ref{Fig:Kleopatra} shows hypothetical satellites of Kleopatra, it can be used a test case for a satellite around any contact binary with a similar shape to Kleopatra, such as Arrokoth. \citet{Richardson2019} compiled a list of 32 Main Belt asteroids, near-Earth asteroids, and comet nuclei with well constrained shape models (see Table 1 in their paper). The most extreme aspect ratio objects with $c/a<0.4$ (similar to that of Kleopatra where $c/a\sim 0.37$) include (4179) Toutatis ($c/a\sim 0.4$), (25143) Itokawa ($c/a\sim 0.39$) (visited by the Hayabusa spacecraft \citep{Fujiwara2006}), (8567) 1996 HW1 ($c/a\sim 0.39$) \citep{Magri2011}, (433) Eros ($c/a\sim 0.33$) ( visited by  the NEAR-Shoemaker spacecraft in 1999-2001 \citep{Robinson2002}), the comet nucleus (103P) Hartley 2 ($c/a\sim 0.31$)\citep{Thomas2013}, (243) Ida ($c/a\sim 0.28$) (visited by the Gallileo spacecraft \citep{Thomas1996}), and (1620) Geographos ($c/a\sim 0.36$) \citep{Hudson1999}.  \citet{Ryabova2002} found that the rotational motion of (1620) Geographos was not enough to cause the material throw-off from the asteroid’s surface. The existence of these extreme aspect ratio bodies implies that there exist a large population of contact binaries and irregularly shaped bodies in the Solar System. Therefore, future surveys should find many more similarly deformed objects.  If any of these known objects or similarly aspherical detected objects have small satellites, they could feasibly  exhibit chaotic rotation from the overlap of the spin-spin and spin-orbit resonances. Indeed, even if the probability of spin-spin interactions is small, the sheer number of readily-observable contact binaries in the Solar System may make chaotic spin-spin coupling a wide-spread phenomenon.

Beyond the Solar System, we note that the formalism developed herein may have applications to stellar dynamics. To this end, \citet{Tremaine2020} pointed out that quadruple star systems consisting of two compact binaries exhibited an overabundance of orbit-orbit resonances, specifically at the 3:2 and 1:1 orbit resonances. The two dumbbell concentric mass model could be especially pertinent to such quadruple stellar systems which, as long as the orbits do not evolve quickly compared to dynamical timescales, are effectively all quadrupole moments, where $M_P$ and $M_S=0$. The orbit of each binary is analogous to the spin of a single body with a permanent quadrupole moment. Intriguingly, the lack of a central mass in the primary would result in effective  ``spin-spin"
resonances having comparable strengths to the spin-orbit resonances in these systems. Therefore, these stellar systems may also exhibit interesting dynamics, akin to spin-spin coupling described in this work. 

\section{Acknowledgements}
 We thank Kaitlin Kratter, Michele Bannister, Robert Jedicke, Edwin Kite, Cristobal Petrovich  and  Mahdi Jafari Nadoushan  for useful conversations.  We thank the anonymous referee for a constructive and insightful report, and for encouraging us to explore the effects of obliquity, which greatly strengthened our manuscript. 
\bibliography{sample63}{}
\bibliographystyle{aasjournal}



\end{document}